\documentclass[useAMS,usenatbib]{mn2e}

\usepackage{amssymb,epsfig,color}
\usepackage{amsmath}
\usepackage{graphicx}
\graphicspath{{figures/}}

\usepackage{siunitx}
\usepackage{booktabs}
\usepackage{multirow}
\DeclareSIUnit\arcsecond{arcsec}
\DeclareSIUnit\parsec{pc}
\DeclareSIUnit\milliarcsecond{mas}
\DeclareSIUnit\year{yr}

\usepackage[hyperindex,breaklinks=true, colorlinks,citecolor=blue,draft=False]{hyperref}  
 \usepackage{flushend} 

\def\reff@jnl#1{{\rm#1\/}}

\def\aj{\reff@jnl{AJ}}                  
\def\araa{\reff@jnl{ARA\&A}}            
\def\apj{\reff@jnl{ApJ}}                
\def\apjl{\reff@jnl{ApJ}}               
\def\apjs{\reff@jnl{ApJS}}              
\def\ao{\reff@jnl{Appl.Optics}}         
\def\apss{\reff@jnl{Ap\&SS}}            
\def\aap{\reff@jnl{A\&A}}               
\def\aapr{\reff@jnl{A\&A\simRev.}}         
\def\aaps{\reff@jnl{A\&AS}}             
\def\azh{\reff@jnl{AZh}}                        
\def\baas{\reff@jnl{BAAS}}              
\def\jrasc{\reff@jnl{JRASC}}            
\def\memras{\reff@jnl{MmRAS}}           
\def\mnras{\reff@jnl{MNRAS}}            
\def\nar{\reff@jnl{New Astronomy Reviews}}            
\def\pra{\reff@jnl{Phys.Rev.A}}         
\def\prb{\reff@jnl{Phys.Rev.B}}         
\def\prc{\reff@jnl{Phys.Rev.C}}         
\def\prd{\reff@jnl{Phys.Rev.D}}         
\def\prl{\reff@jnl{Phys.Rev.Lett}}      
\def\pasa{\reff@jnl{PASA}}              
\def\pasp{\reff@jnl{PASP}}              
\def\pasj{\reff@jnl{PASJ}}              
\def\qjras{\reff@jnl{QJRAS}}            
\def\skytel{\reff@jnl{S\&T}}            
\def\solphys{\reff@jnl{Solar\simPhys.}}    
\def\sovast{\reff@jnl{Soviet\simAst.}}     
 \def\ssr{\reff@jnl{Space\simSci.Rev.}}    
\def\zap{\reff@jnl{ZAp}}                
\def\nat{\reff@jnl{Nature}}             

\title[AME Spectral Variations with \mbox{QUIJOTE} and \mbox{C-BASS}]{Detection of Spectral Variations of Anomalous Microwave Emission with \mbox{QUIJOTE} and \mbox{C-BASS}}
\author[R. Cepeda-Arroita et al.]{R.\,Cepeda-Arroita,$\!^{1}$\thanks{\url{E-mail: roke.cepeda-arroita@manchester.ac.uk}} S.\,E.\,Harper,$\!^{1}$ C.\,Dickinson,$\!^{1,2}$ J.\,A.\,Rubi\~no-Mart\'{i}n,$\!^{3,4}$ \newauthor R.\,T.\,G\'{e}nova-Santos,$\!^{3,4}$ Angela\,C.\,Taylor,$\!^{5}$ T.\,J.\,Pearson,$\!^{2}$ M.\,Ashdown,$\!^{6,7}$ \newauthor A.\,Barr,$\!^{1}$ R.\,B.\,Barreiro,$\!^{8}$ B.\,Casaponsa,$\!^{8}$ F.\,J.\,Casas,$\!^{8}$ H.\,C.\,Chiang,$\!^{9,10}$ \newauthor R.\,Fernandez-Cobos,$\!^{8}$ R.\,D.\,P.\,Grumitt,$\!^{5}$ F.\,Guidi,$\!^{3,4}$ H.\,M.\,Heilgendorff,$\!^{10}$ \newauthor D.\,Herranz,$\!^{8}$ L.\,R.\,P.\,Jew,$\!^{5}$ J.\,L.\,Jonas,$\!^{11,12}$ Michael\,E.\,Jones,$\!^{5}$ A.\,Lasenby,$\!^{6,7}$ \newauthor J. Leech,$\!^{5}$ J.\,P.\,Leahy,$\!^{1}$ E.\,Mart\'{i}nez-Gonz\'{a}lez,$\!^{8}$ M.\,W.\,Peel,$\!^{1,3,4}$ L.\,Piccirillo,$\!^{1}$ \newauthor F.\,Poidevin,$\!^{3,4}$ A.\,C.\,S.\,Readhead,$\!^{2}$ R.\,Rebolo,$\!^{3,4,13}$ B.\,Ruiz-Granados,$\!^{3,4,14,15}$ \newauthor J.\,Sievers,$\!^{9,16}$ F.\,Vansyngel,$\!^{3,4}$ P.\,Vielva,$\!^{8}$ R.\,A.\,Watson,$\!^{1}$ \\
$^{1}$Jodrell Bank Centre for Astrophysics, Alan Turing building, Department of Physics and Astronomy, School of Natural \\
Sciences, The University of Manchester, Oxford Road, Manchester, M13 9PL, Manchester, U.K. \\
$^{2}$Cahill Centre for Astronomy and Astrophysics, California Institute of Technology, Pasadena, CA 91125, USA \\
$^{3}$Instituto de Astrof\'{i}sica de Canarias, 38200 La Laguna, Tenerife, Canary Islands, Spain \\
$^{4}$Departamento de Astrof\'{i}sica, Universidad de La Laguna (ULL), 38206 La Laguna, Tenerife, Spain \\
$^{5}$Sub-department of Astrophysics, University of Oxford, Denys Wilkinson Building, Keble Road, Oxford OX1 3RH, UK \\
$^{6}$Astrophysics Group, Cavendish Laboratory, University of Cambridge, J.J. Thomson Avenue, Cambridge CB3 0HE, UK \\
$^{7}$Kavli Institute for Cosmology, University of Cambridge, Madingley Road, Cambridge CB3 0HA, UK \\
$^{8}$Instituto de F\'{i}sica de Cantabria (CSIC-Universidad de Cantabria), Avda. de los Castros s/n, 39005 Santander, Spain \\
$^{9}$Department of Physics, McGill University, 3600 Rue University, Montr\'{e}al, QC H3A 2T8, Canada \\
$^{10}$School of Mathematics, Statistics \& Computer Science, University of KwaZulu-Natal, Westville Campus, \\ Private Bag X54001, Durban 4000, South Africa \\
$^{11}$Department of Physics and Electronics, Rhodes University, Grahamstown, 6139, South Africa \\
$^{12}$South African Radio Astronomy Observatory, 2 Fir Road, Observatory, Cape Town, 7925, South Africa \\
$^{13}$Consejo Superior de Investigaciones Cient\'{i}ficas, Spain \\
$^{14}$Space Science Data Center - Agenzia Spaziale Italiana, Via del 
Politecnico snc, 00133, Rome, Italy \\
$^{15}$Istituto Nazionale di Fisica Nucleare, Sezione di Roma 2, Via 
della Ricerca Scientifica 1, 00133 Rome, Italy \\
$^{16}$School of Chemistry \& Physics, University of KwaZulu-Natal, Westville Campus, \\ Private Bag X54001, Durban 4000, South Africa
}

\begin{document}

\date{Accepted XXX. Received YYY; in original form ZZZ}

\pagerange{\pageref{firstpage}--\pageref{lastpage}} \pubyear{2021}

\maketitle

\label{firstpage}


\begin{abstract}
\noindent Anomalous Microwave Emission (AME) is a significant component of Galactic diffuse emission in the frequency range $10$--$60\,\si{\giga\hertz}$ and a new window into the properties of sub-nanometre-sized grains in the interstellar medium. We investigate the morphology of AME in the $\approx\ang{10}$ diameter \mbox{$\lambda$ Orionis} ring by combining intensity data from the \mbox{QUIJOTE} experiment at $11$, $13$, $17$ and $19\,\si{\giga\hertz}$ and the C-Band All Sky Survey (\mbox{C-BASS}) at $4.76\,\si{\giga\hertz}$, together with 19 ancillary datasets between $1.42$ and $3000\,\si{\giga\hertz}$. Maps of physical parameters at $\ang{1}$ resolution are produced through Markov Chain Monte Carlo (MCMC) fits of spectral energy distributions (SEDs), approximating the AME component with a log-normal distribution. AME is detected in excess of $20\,\sigma$ at degree-scales around the entirety of the ring along photodissociation regions (PDRs), with three primary bright regions containing dark clouds. A radial decrease is observed in the AME peak frequency from $\approx35\,\si{\giga\hertz}$ near the free-free region to $\approx21\,\si{\giga\hertz}$ in the outer regions of the ring, which is the first detection of AME spectral variations across a single region. A strong correlation between AME peak frequency, emission measure and dust temperature is an indication for the dependence of the AME peak frequency on the local radiation field. The AME amplitude  normalized by the optical depth is also strongly correlated with the radiation field, giving an overall picture consistent with spinning dust where the local radiation field plays a key role.
\end{abstract}



\section{Introduction}

\subsection{Anomalous Microwave Emission}

Anomalous Microwave Emission (AME) is a major component of Galactic diffuse emission between $10 < \nu < 60\,\si{\giga\Hz}$, discovered by an excess of emission strongly correlated with far-infrared (FIR) emission that could not be explained by synchrotron or free-free emission \citep{Leitch1997, Kogut1996, Oliveira1998}. We now know this correlation exists down to $2\,$arcmin scales \citep{Casassus2006, Casassus2008, Scaife2009, Dickinson2009a, Dickinson2010, Tibbs2013, Battistelli2019, Arce2019} with the caveat that most of the emission is diffuse at larger angular scales, and that AME is generally associated with colder dust in the range $14$--$20\,\si{\K}$ \citep{Planck2014_AME}. Extragalactic detections of AME have also been made, where AME has been associated with mid-infrared (MIR) counterparts \citep{Murphy2010, Murphy2018}. The prevailing idea is that the majority of AME arises from electric dipole emission from rapidly rotating dust grains \citep{Draine1998a}. This spinning dust hypothesis is increasingly favoured by observational results, although alternative mechanisms have been proposed \citep{Draine1999, Bennett2003b, Meny2007, Jones2009, Nashimoto2019, Nashimoto2020}. For the purpose of this paper, it is assumed that the majority of AME originates from spinning dust grains. Observations suggest that AME is ubiquitous along the Galactic plane and that it accounts for up to half of the flux density in intensity at $\sim 30\,\si{\giga\Hz}$ \citep{Planck2014_AME, Planck2015_XXV}, being strongest in photodissociation regions (PDRs) such as Perseus \citep{Watson2005, Tibbs2010, Planck2011_XX}, $\rho$ Ophiuchi \mbox{\citep{Casassus2008, Planck2011_XX}}, molecular clouds \citep{Planck2014_AME, Poidevin2019, Genova-Santos2011, Planck2013_XII} and other dense environments associated with H\textsc{ii} regions \mbox{\citep{Todorovic2010}}. A comprehensive review of AME is given in \cite{Dickinson2018AMEReview}.

\subsection{Spinning Dust}
The electric dipole spinning dust mechanism was first proposed by \cite{Erickson1957}. This possibility was pursued further due to the high degree of correlation between AME and far-infrared (FIR) emission from larger grains observed by CMB experiments in the late 1990s \citep{Leitch1997,Kogut1996,Oliveira1998}, prompting a spinning dust theory by \cite{Draine1998b}. This model describes the exchange of angular momentum between dust grains and their environment, including collisions and the absorption and emission of photons. In grains with an electric dipole moment that is misaligned with the axis of rotation, electric dipole emission is generated at the rotational frequency. The total power emitted is dominated by the fastest spinning grains, which are in turn the smallest. This is the reason naturally abundant nanoparticles such as polycyclic aromatic hydrocarbons or PAHs \citep{Draine1998a}, nanosilicates \citep{brandon_no_pahs, Hensley2017}, fullerenes \citep{Iglesias-Groth2005} and nanodiamonds \citep{Greaves2018} have been proposed as being dominant spinning dust carriers. Despite the rise in emitted power with frequency, few grains achieve very high-frequencies due to the intrinsic thermal cut-off since the fastest spinning grains lose their energy more rapidly and the fact that strong radiation fields can break the dust grains and ionize the interstellar medium, effectively reducing spinning dust emission. It is this trade-off that gives rise to the peaked spectrum observed.

Despite the simplicity of the single dipole case, the shape of the spectrum depends on the distribution of grain sizes, dipole moments, densities and the local radiation field, making it difficult to model. Several improvements to the \cite{Draine1998a} model have been made by \cite{Ali-Hamoud2009, Ysard2010a, Hoang2010, Hoang2011} and \cite{Silsbee2011}. The most recent spinning dust modelling code, which integrates some of these changes, is the \textsc{SpDust2}\,\footnote{\url{http://pages.jh.edu/~yalihai1/spdust/spdust.html}} model \citep{Ali-Haimoud2010, Silsbee2011}. The observed AME distribution is often a superposition of components, resulting in a high-dimensional parameter space that makes linking the shape of the observed AME spectrum with theory challenging without $\sim1\,\si{\giga\hertz}$ spectral resolution data points in the range $10$--$60\,\si{\giga\hertz}$ and more frequency points than are currently available. Conversely, a robust understanding of the theory would provide a direct way of measuring the many astrophysical parameters on which spinning dust emission depends.

One of the strongest observational arguments for the spinning dust hypothesis as the dominant emission mechanism of AME is the lack of observed polarization, since many alternative mechanisms imply measurable polarization fractions. Upper limits on the polarization fraction of the order of 1 per cent on $\ang{1}$ scales \citep{Dickinson2011, Lopez-Caraballo2011, Rubino-Martin2012a, Genova-Santos2017} and upper limits of a few percent at arcminute scales \citep{Mason2009, Battistelli2015} have been set. The current absence of AME polarization detections also highlights the importance of intensity data as a bridge between observations and theory, and in turn, a tool to understand the level at which polarized emission is expected, which is important for both astrophysics and CMB polarization observations.

\subsection{The \mbox{$\lambda$ Orionis} Ring}
The $\approx\ang{10}$ diameter \mbox{$\lambda$ Orionis} ring is a dense neutral hydrogen shell surrounding the expanding H\textsc{ii} region \mbox{Sh2-264} \citep{Sharpless1959}, which is ionized by O8 III star \mbox{$\lambda$ Orionis} and its B-type associates, roughly $5\,\si{\mega\year}$ old \citep{Murdin1977}. Observationally, the large angular extent of \mbox{$\lambda$ Orionis} makes it a good case study to test the spinning dust hypothesis through the direct detection of AME using degree-resolution data, with the advantages that it has an approximately circular symmetry and a clear separation between the H\textsc{ii} bubble and the surrounding dust. The H\textsc{ii} region, a good example of a Str\"{o}mgren sphere, lies at a distance of $\sim420\,\si{\parsec}$ \citep{Schlafly_2014}, implying the ring is $\sim70\,\si{\parsec}$ across. Surrounding the free-free emission-dominated H\textsc{ii} shell is a ring of dust, which can be seen through thermal dust emission above $\sim100\,\si{\giga\hertz}$. The interface between the colder molecular clouds and the ionized H\textsc{ii} region near the inner edge of the ring is a photodissociation region, which is a favourable environment for bright AME due to its high density and radiation field \citep{Dickinson2018AMEReview}. The PDR, starting at a radial distance of $\approx4^{\circ}$ from star \mbox{$\lambda$ Orionis}, hosts an environment in which small grains can be charged by and simultaneously protected from the ionizing radiation.

\begin{figure*}
	\begin{center}
		\includegraphics[width=\textwidth,angle=0]{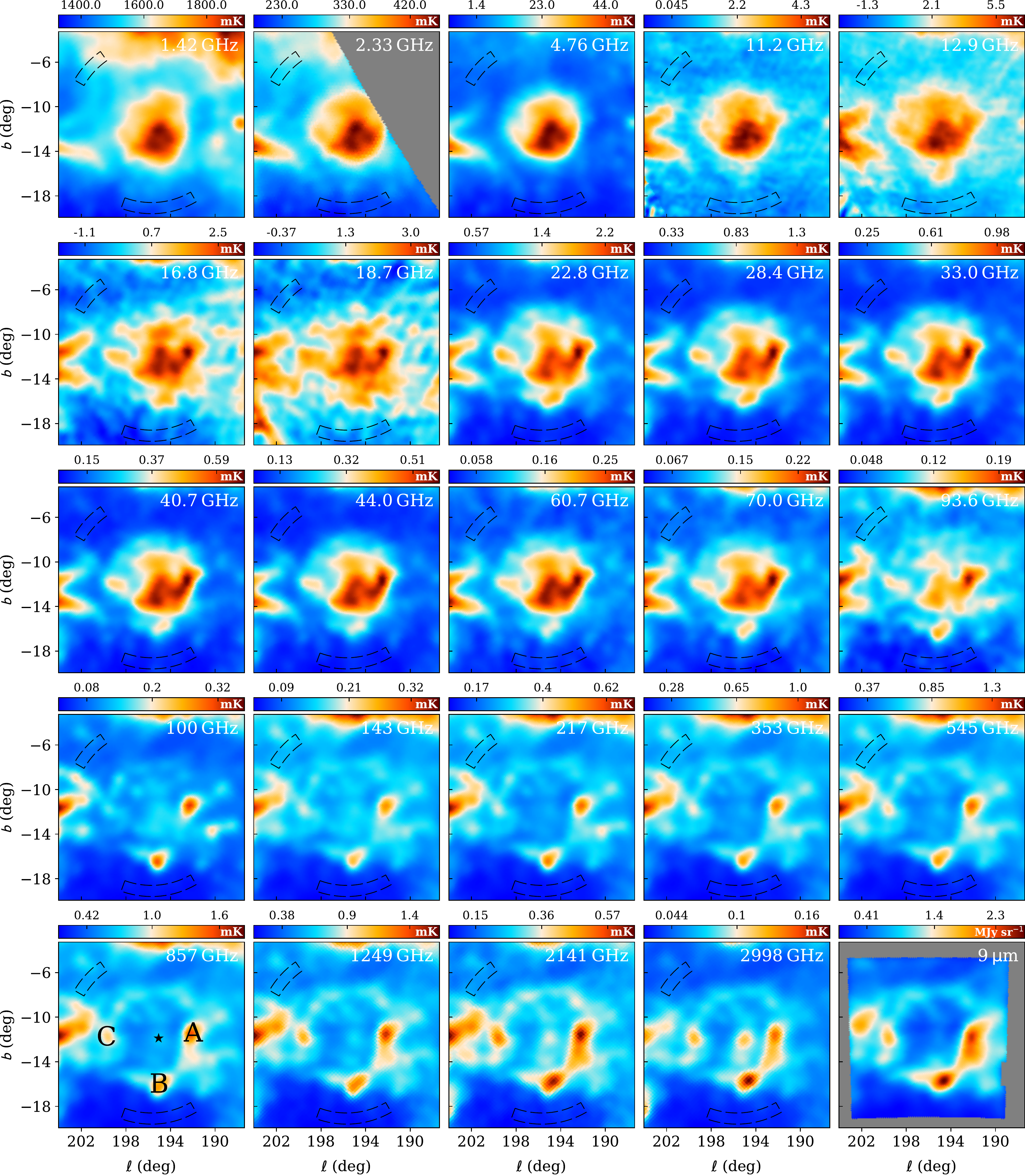}
		\caption{Multi-frequency view of \mbox{$\lambda$ Orionis} in Galactic coordinates centred on G195.7$-$11.6 and spanning $16.\!^{\circ}7$ on a side, corresponding to frequencies in Table\,\ref{tab:surveys}. All maps are smoothed to a FWHM of $\ang{1}$. Maps are also CMB-subtracted above $10\,\si{\GHz}$ and shown in $\si{\milli\kelvin}$, with the exception of the PAH emission-dominated AKARI $9\,\si{\micro\meter}$ map. The backgrounds used for aperture photometry are delimited by the dashed lines. The colour scale ticks are set to exactly match the values shown. Key regions \textit{A} (B30), \textit{B} (B233, LDN1588, LDN1589, LDN1590) and \textit{C} (LDN1602 and LDN1603) are shown in the \textit{Planck} $857\,\si{\giga\hertz}$ map, where the position of O8 III star \mbox{$\lambda$ Orionis} is indicated by the star-shaped marker. AME in all three regions is evident at \mbox{QUIJOTE} frequencies and up to $\sim30\,\si{\giga\hertz}$. Grey areas are where no data are available. Note that the declination band seen most clearly in the QUIJOTE 18.7 GHz map corresponds to the geostationary satellite band, and is well away from the region analyzed.}
		\label{fig:multifrequency}
	\end{center}
\end{figure*}

While the ring has not been widely studied as a whole due to its large angular scale, it has been observed in H\textsc{i} \citep{Wade1957, Zhang1991}, CO and FIR. CO observations by \cite{Maddalena1987} suggest that the ring was formed by an initially flattened molecular cloud that expanded into a shell, primarily disrupted by \mbox{$\lambda$ Orionis}, projecting the current ring shape. Far-infrared IRAS data \mbox{\citep{Zhang1989}} revealed an almost perfectly symmetric dust ring with a fragmented substructure, heated by a strong radiation field dominated by star \mbox{$\lambda$ Orionis}, whose position is shown in Fig.\,\ref{fig:multifrequency}. More recent CO observations \citep{Lang2000,Lang1998} and simulations of the ring's morphology \citep{Lee_2015} support the basic shell model with a secondary toroidal ring component, favouring the flattened progenitor molecular cloud hypothesis. The star \mbox{$\lambda$ Orionis}, which dominates UV emission in the region, has a proper motion of $\approx3\,\si{\milliarcsecond}\,\si{\year}^{-1}$ \citep{Hipparcos2007} and is offset about a degree from the geometrical centre of the ring \citep{Dolan2002}.

The region is shown in Fig.\,\ref{fig:multifrequency}. At a FWHM of $1^{\circ}$ it displays three distinct cores labelled \textit{A}, \textit{B} and \textit{C}, each consisting of one or several dark and molecular clouds \citep{Barnard1927,Lynds1962}, with many smaller clouds lying around the ring. While region \textit{A} is dominated by dark cloud B30, region \textit{B} is a mixture of dark clouds B223, LDN1588, LDN1589 and LDN1590. Likewise, region \textit{C} consists primarily of LDN1602 and LDN1603. The \mbox{$\lambda$ Orionis} ring is also in the vicinity of the Orion Molecular Cloud Complex, which can be seen a few degrees west of region \textit{C}. The changes in the morphology of the region as a function of frequency can be seen in Fig.\,\ref{fig:multifrequency}, with a free-free emission bubble transitioning into a ring of thermal dust emission above $\sim100\,\si{\giga\hertz}$. An excess of emission above $11\,\si{\giga\hertz}$ can be seen in all three major cores, hinting that the regions are AME-dominated at such frequencies. Due to the higher radiation field on the inner side of the PDR, the highest spinning frequencies are also expected on the inner side, where the stronger radiation can break grains such as PAHs into smaller grain populations \citep{Tielens2008} and impart greater angular momentum. The ring was first identified as an interesting AME source by the \cite{Planck2015_XXV} due to the high degree of correlation between AME and thermal dust emission near region \textit{A}.

The data used are introduced in Section \ref{sec:obs}, followed by the emission models and methods in Section \ref{sec:sed_fitting}. Results and their physical interpretation are presented in Section \ref{sec:discussion}.


\section{Maps}
\label{sec:obs}

\subsection{QUIJOTE}

The \mbox{QUIJOTE} (Q-U-I JOint TEnerife) experiment \citep{Genova-Santos2015a, QUIJOTE2017} uses a set of microwave polarimeters at the Teide Observatory in the Canary Islands, $2400\,\si{\meter}$ above sea level. It consists of two $2.25\,\si{\meter}$ crossed-Dragone telescopes and three instruments, spanning the frequency range $10$--$40\,\si{\giga\hertz}$ with angular resolutions ranging from $\ang{0.9}$ at the lowest frequency to $\ang{0.3}$ at the highest one. Here we use data from the Multi-Frequency Instrument (MFI), which started operations in November 2012. MFI operates at $11$, $13$, $17$ and $19\,\si{\giga\hertz}$, which is an ideal range to detect the low-frequency upturn of AME. In particular, intensity data in \mbox{$\lambda$ Orionis} from the \mbox{QUIJOTE} wide survey smoothed to a FWHM of $\ang{1}$ are used, with a total of $\approx8500$ hours at each frequency covering $\approx70$ per cent of the full sky over 5 years of observations. \mbox{QUIJOTE} data enable the measurement of the low-frequency tail of AME, and together with \mbox{\textit{WMAP}} and \textit{Planck} maps between $22.8 < \nu < 60.8\,\si{\giga\hertz}$, a full measurement of the AME curve can be obtained. This allows us to directly measure the AME peak frequency and width with more accuracy than in \cite{Planck2014_AME} without assuming priors on the width or using widths from single-component \textsc{SpDust} models. The calibration uncertainties range from $3$ per cent at the lowest two frequencies to $5$ per cent at the highest two frequencies. For detailed characteristics, operation and noise properties see Rubi\~no-Mart\'{i}n et al. (in prep.).

\subsection{C-Band All Sky Survey}

The \mbox{C-BASS} project \citep{project_paper} is an experiment to map the full sky at an effective frequency of $4.76\,\si{\giga\hertz}$ and an angular resolution of $\approx45\,$arcmin. We use data from the northern \mbox{C-BASS} telescope located in Owens Valley Radio Observatory, California, which uses a $6.1\,\si{\meter}$ Gregorian antenna. It covers the band $4.5$--$5.5\,\si{\giga\hertz}$ and has mapped the northern sky above a declination of $-15.\!^{\circ}6$ for an effective observing time of $\approx2200$ hours between November 2012 and March 2015. While the astronomical calibration uncertainty is of the order of $1$ per cent (Pearson et al. in prep. and Taylor et al. in prep.), an absolute calibration uncertainty of $3$ per cent is adopted in this paper. This is primarily to account for uncertainties in the scale calibration coming from uncertainties in the shape of the beam and sidelobes used for deconvolution. In future data releases, an absolute uncertainty of $\approx1$ per cent is expected. \mbox{C-BASS} enables us to measure the level of free-free emission in the region much more accurately than with older surveys below $5\,\si{\giga\hertz}$, and in turn plays a major role in constraining the amplitude of the AME. We use the deconvolved intensity map presented in the upcoming northern survey paper, smoothed to a FWHM of $\ang{1}$. Full details of the \mbox{C-BASS} northern survey are given in Pearson et al. (in prep.) and Taylor et al. (in prep.).

\subsection{Ancillary Data}

Nineteen ancillary datasets between $1.42$ and $3000\,\si{\giga\hertz}$ are used in conjunction to \mbox{QUIJOTE} and \mbox{C-BASS} data, listed in Table\,\ref{tab:surveys} and mapped in Fig.\,\ref{fig:multifrequency}. These are also smoothed to a FWHM of \ang{1} and a smoothed CMB estimate is subtracted at frequencies above $10\,\si{\giga\hertz}$. In particular, we use the Needlet Internal Linear Combination (\textsc{NILC}) CMB map \citep{Planck2018_IV}. Other CMB separation methods such as \textsc{SMICA}, \textsc{SEVEM} and \textsc{Commander} were also tested, finding no significant differences in our analysis and results. This is expected since differences in the region at $28.4\,\si{\giga\hertz}$ near the peak frequency of the AME are of the order of 0.5 per cent.

The lowest frequency survey at $1.420\,\si{\giga\hertz}$, which is calibrated to within $5$ per cent \citep{Reich1986}, is assigned an effective calibration uncertainty to account for the unknown full-beam to main-beam calibration of the dataset. A preliminary analysis by \cite{Irfan2014} suggests that the scale calibration between extended and degree scales can be up to a factor of $\sim2$, differing from the $1.55\pm0.08$ factor between $1^{\circ}$ and $7^{\circ}$ in \cite{Reich1988}. Therefore, we calibrated the dataset to degree-scales using the official factor and a $30$ per cent overall calibration uncertainty is assigned to account for the brightness temperature scale not being constant with angular scale, as shown in Table\,\ref{tab:surveys}. However, given the large calibration uncertainties in Reich, the dataset does not contribute as much as the \mbox{C-BASS} and \mbox{HartRAO} maps, and acts primarily as a check. The \mbox{HartRAO} map by \cite{Jonas1998} is also used despite it not covering the north-eastern edge of the \mbox{$\lambda$ Orionis} ring (see Fig.\,\ref{fig:multifrequency}), and measurements are only considered one aperture away from this edge. Together with the \mbox{C-BASS} map, these two maps constrain the free-free emission level in the region.

In conjunction, we use \textit{WMAP} \mbox{\citep{Bennett2013}}, \textit{Planck} \citep{Planck2018_I} and COBE-DIRBE \citep{Hauser1998} data up to $3\,\si{\tera\hertz}$, while avoiding molecular CO lines by excluding the $100\,\si{\giga\hertz}$ and $217\,\si{\giga\hertz}$ \textit{Planck} HFI frequencies from the SED fit \mbox{\citep{Planck2013_CO}}. \textit{WMAP} and \textit{Planck} data above $100\,\si{\giga\hertz}$ primarily measure the thermal dust emission curve, while data below $100\,\si{\giga\hertz}$ constrain the high-frequency end of AME.

We also use the AKARI $9\,\si{\micro\meter}$ data of the region used in \cite{Bell2019}. While this dataset is not used for SED analysis, it is used to assess correlations with AME in Section\,\ref{sec:AKARI}. The band covers the range $6<\lambda<12\,\si{\micro\meter}$, covering PAH-emission peaks at 6.2, 7.7, 8.6 and $11.2\,\si{\micro\meter}$. This map has an original resolution of $\sim10\,\si{\arcsecond}$, which allows us to remove stellar emission by using a median filter to remove the well-separated small-scale bright distributions of stars from the diffuse emission before smoothing the maps to a FWHM of $\ang{1}$, as shown in Fig.\,\ref{fig:multifrequency}.

\begin{table*}
\caption{Summary of multi-frequency data, where $\nu$ is the effective frequency, $\Delta\nu$ denotes bandwidth, $\delta$ is declination in \mbox{J2000} celestial coordinates and $\sigma_{\rm cal}$ is the effective calibration uncertainty. The effective calibration uncertainty accounts for scale-calibration uncertainties due to beam inefficiencies, which primarily affect the $1.42\,\protect\si{\giga\hertz}$ map.\,$^{\dagger}\,$The AKARI $9\,\si{\micro\meter}$ dataset, provided by T. Onaka and A. Bell, only covers the $\lambda$ Orionis region and is not used in SED fitting. \label{tab:surveys}}
\resizebox{\textwidth}{!}{
\begin{tabular}{ccccccccc}
\toprule
	\multirow{ 2}{*}{\textbf{Telescope}} & \boldmath{$\nu$} & \boldmath{$\Delta\nu$} & \textbf{FWHM} & \textbf{Sky} & \boldmath{$\sigma_{\rm cal}$} & \multirow{ 2}{*}{\textbf{Reference}} \\
	& \textbf{(GHz)} & \textbf{(GHz)} & \textbf{(arcmin)} & \textbf{Coverage} & \textbf{(\%)} & \\  \midrule
	Stockert/Villa-Elisa & 1.420 & 0.014 & 35 & Full Sky & 30 & \cite{Reich2001} \\
	HartRAO & 2.326 & 0.040 & 20 & $\ang{-83}<\delta<12.\!^{\circ}5$ & 10 & \cite{Jonas1998} \\
	\mbox{C-BASS} & 4.76 & 0.49 & 44 & $\delta>-15.\!^{\circ}6$ & 3 & Taylor et al. (in prep.) \\
	\mbox{QUIJOTE} MFI & 11.2 & 2 & 56 & $\delta\gtrsim\ang{0}$ & 3 & Rubi\~no-Mart\'{i}n et al. (in prep.)\\
	\mbox{QUIJOTE} MFI & 12.9 & 2 & 56 & $\delta\gtrsim\ang{0}$ & 3 & Rubi\~no-Mart\'{i}n et al. (in prep.)\\
	\mbox{QUIJOTE} MFI & 16.8 & 2 & 38 & $\delta\gtrsim\ang{0}$ & 5 & Rubi\~no-Mart\'{i}n et al. (in prep.)\\
	\mbox{QUIJOTE} MFI & 18.7 & 2 & 38 & $\delta\gtrsim\ang{0}$ & 5 & Rubi\~no-Mart\'{i}n et al. (in prep.)\\
	\textit{WMAP} K & 22.8 & 5.5 & 49 & Full Sky & 1 & \cite{WMAP2013_RESULTS} \\
	\textit{Planck} LFI & 28.4 & 6 & 32.4 & Full Sky & 1 & \cite{Planck2018_I} \\
	\textit{WMAP} Ka & 33 & 7 & 40 & Full Sky & 1 & \cite{WMAP2013_RESULTS} \\
	\textit{WMAP} Q & 40.7 & 8.3 & 31 & Full Sky & 1 & \cite{WMAP2013_RESULTS} \\
	\textit{Planck} LFI & 44.1 & 8.8 & 27.1 & Full Sky & 1 & \cite{Planck2018_I} \\
	\textit{WMAP} V & 60.8 & 14 & 21 & Full Sky & 1 & \cite{WMAP2013_RESULTS} \\
	\textit{Planck} LFI & 70.4 & 14 & 13.3 & Full Sky & 1 & \cite{Planck2018_I} \\
	\textit{WMAP} W & 93.5 & 20.5 & 13 & Full Sky & 1 & \cite{WMAP2013_RESULTS} \\
	\textit{Planck} HFI & 100 & 33 & 9.7 & Full Sky & 1 & \cite{Planck2018_I} \\
	\textit{Planck} HFI & 143 & 47 & 7.3 & Full Sky & 1 & \cite{Planck2018_I} \\
	\textit{Planck} HFI & 217 & 72 & 5.0 & Full Sky & 1 & \cite{Planck2018_I} \\
	\textit{Planck} HFI & 353 & 100 & 4.8 & Full Sky & 1.3 & \cite{Planck2018_I} \\
	\textit{Planck} HFI & 545 & 180 & 4.7 & Full Sky & 6.0 & \cite{Planck2018_I} \\
	\textit{Planck} HFI & 857 & 283 & 4.3 & Full Sky & 6.4 & \cite{Planck2018_I} \\
	COBE-DIRBE 240 & 1249 & 495 & 42 & Full Sky & 13.5 & \cite{Hauser1998} \\
	COBE-DIRBE 140 & 2141 & 605 & 38 & Full Sky & 10.6 & \cite{Hauser1998} \\
	COBE-DIRBE 100 & 2998 & 974 & 39 & Full Sky & 11.6 & \cite{Hauser1998} \\ 
	AKARI$\,^{\dagger}$ & $9\,\si{\micro\meter}$ & $\sim 6\,\si{\micro\meter}$ & $\sim0.17$ & Full Sky & $<10\%$ & \cite{Bell2019} \\ 
	\bottomrule
\end{tabular}}
\end{table*}


\section{SED Fitting}
\label{sec:sed_fitting}

\subsection{Foreground Modelling}
\label{sec:foreground_modelling}

Foregrounds around the \mbox{$\lambda$ Orionis} ring are modelled using the superposition of free-free, AME and thermal dust models, resulting in an overall SED with 7 free parameters:
\begin{equation}
\begin{split}
S_{\rm total}(\nu) = S_{\rm ff}(\nu, \mathrm{EM}) + S_{\rm d}(\nu, \beta, T_{\rm d}, \tau_{\rm 353}) \\ + S_{\rm AME}(\nu, A_{\rm AME}, \nu_{\rm AME}, W_{\rm AME})\,,
\end{split}
\end{equation}
where $S_{\rm ff}$, $S_{\rm d}$ and $S_{\rm AME}$ are the free-free, thermal dust and AME contributions, respectively, and the 7 free parameters are defined in Sections \ref{sec:free-free}--\ref{sec:thermal_dust}. Models and observables are described in the following subsections. Synchrotron emission is not modelled since the H\textsc{ii} region is dominated by free-free emission at the frequencies used. The synchrotron component is smooth and relatively low across the region, contributing to less than 5 per cent of the total emission in the H\textsc{ii} bubble at $1.42\,\si{\giga\hertz}$ with the background subtraction used in SED measurements, as estimated by the \cite{Planck2015_X} \textsc{Commander} separation. Therefore, we do not use the 408\,MHz map \citep{Haslam1982} due to the fact that using this and fitting a synchrotron model would add degeneracies at low frequencies, since we would only be able to fit a single synchrotron parameter, either the amplitude or spectral index, with effectively a single data point. The effective calibration uncertainties of the \cite{Haslam1982} map, particularly due to the unknown main beam to full beam calibration, are $\gtrsim10$ per cent \citep{Remazeilles2015}. The calibration uncertainties for low frequency surveys up to \mbox{C-BASS} are also larger than the expected synchrotron contribution, making these points unsuitable to separate free-free and any residual synchrotron on their own. The overall model is justified by the lack of evidence of a steep synchrotron component in the fitted spectra or in spectral indices of the region.

\subsubsection{Free-free Emission}
\label{sec:free-free}

Free-free emission is mostly seen in the H\textsc{ii} region Sh-2-264 inside the \mbox{$\lambda$ Orionis} ring. The flux density is given by
\begin{equation}
S_{\rm ff}(\nu) = \frac{2 k_{\rm B} \nu^{2}}{c^{2}} \cdot \Omega_{\rm b} \cdot T_{\rm ff}(\nu)\,,
\end{equation}
where $\nu$ is the observing frequency, $k_{\rm B}$ is the Boltzmann constant, $\Omega_{\rm b}$ is the source solid angle and $T_{\rm ff}$ is the free-free emission brightness temperature. We use the free-free brightness temperature model in \cite{Draine_book}:
\begin{equation}
\begin{split}
T_{\rm ff}(\nu) = T_{\rm e} \cdot \left \{ 1 - \exp \left [ -\tau_{\rm{ff}}(\nu) \right ] \right \}\,,
\end{split}
\end{equation}
where $T_{\rm e}$ is the electron temperature and $\tau_{\rm{ff}}(\nu)$ is the free-free optical depth, given by
\begin{equation}
\begin{split}
\tau_{\rm ff}(\nu) = 5.468 \cdot 10^{-2} \cdot \rm EM \cdot \textit{T}_{\rm e}^{-\frac{3}{2}} \cdot \left ( \frac{\nu}{GHz} \right )^{-2} \cdot \textit{g}_{\rm ff}(\nu)\,,
\end{split}
\end{equation}
where $\mathrm{EM} \approx \int n_{\rm e}^2\,dl$ is the emission measure along the line of sight, dependent on the electron density $n_e$, and $g_{\rm{ff}}(\nu)$ is the dimensionless free-free Gaunt factor
\begin{equation}
\begin{split}
\begin{aligned}
\noindent \exp \Bigl [ g_{\rm{ff}}(\nu) \Bigl ] =  \\ \exp \Biggl \{ 5.960-\frac{\sqrt{3}}{\pi} \cdot \ln \left [ \frac{\nu}{\si{\giga\hertz}} \left ( \frac{T_{\rm e}}{10^4\,\si{\kelvin}} \right )^{-\frac{3}{2}} \right ] \Biggl \} + \rm{e} \,,
\end{aligned}
\end{split}
\end{equation}
where $\mathrm{e}\approx2.718$ is Euler's number. This model works in both optically thick and thin regimes, although $\lambda$ Orionis is optically thin with typical free-free optical depths of \mbox{$\sim10^{-4}$} at $1\,\si{\giga\hertz}$. The spectrum follows a temperature spectral index of $\beta \approx -2.10$, steepening above $\sim 100\,\si{\giga\Hz}$ to $\beta \approx -2.14$ \citep{Planck2015_XXIII}. Due to the limited dependence of the model on the electron temperature and its narrow range in \mbox{$\lambda$ Orionis}, we use a fixed $T_{\rm e} = 7500\,\si{\K}$, as measured by \cite{Quireza2006} and \cite{Maddalena1987}, leaving $\mathrm{EM}$ as the only free parameter.

\subsubsection{Anomalous Microwave Emission}
\label{sec:AME}

AME is modelled using an empirical log-normal approximation, first proposed by \cite{Stevenson2014_lognormal}. This symmetrical distribution serves as an indicator of the presence, frequency position and width of AME while only introducing 3 free parameters, thus avoiding the high dimensionality and degeneracies of \mbox{{\textsc{SpDust}}} models while making no assumptions about the nature of the emission mechanism:
\begin{equation}
S_{\rm AME}(\nu) = A_{\rm AME} \cdot \exp{\left\{ -\frac{1}{2} \cdot \left[	\frac{ \ln{(\nu/\nu_{\rm AME})}} {W_{\rm AME}} 	  \right]^{2} \right\}}\,,
\end{equation}
\noindent where $A_{\rm AME}$ is the peak amplitude, $\nu_{\rm AME}$ is the peak frequency in flux density, and $W_{\rm AME}$ is the width of the spectrum. The relation between the log-normal and theoretical models is explored by fitting our log-normal model to typical \textsc{SpDust2} models for cold neutral media, dark clouds, molecular clouds, warm ionized media and warm neutral media. We find values between $0.4 < W_{\rm AME} < 0.7$ for these environments, with a mean width of $W_{\rm AME}\approx0.5\pm0.1$. In practice, slightly wider distributions than those in theoretical curves are expected due to the superposition of environments. In all cases, the log-normal models give a very good approximation near the peak of \textsc{SpDust2} models, recovering the original $A_{\rm AME}$ and $\nu_{\rm AME}$ values.

\subsubsection{Thermal Dust Emission}
\label{sec:thermal_dust}

Thermal dust emission, found primarily around the \mbox{$\lambda$ Orionis} ring, is modelled as a single modified black-body curve, given by
\begin{equation}
S_{\rm d}(\nu) = \frac{2 h \nu^3}{c^2} \cdot \frac{	\left( \nu / 353\,\si{\giga\Hz} \right)^{\beta}	}{	e^{h\nu / k_{\rm B} T_{\rm d}} - 1	} \cdot \tau_{\rm 353} \cdot \Omega_b\,,
\end{equation}
\noindent where $\tau_{\rm 353}$ is the optical depth at $353\,\si{\giga\hertz}$, $\beta$ is the emissivity index, and $T_{\rm d}$ is the dust equilibrium temperature \citep{Hildebrand1983, Compiegne2011, Planck2016_XI, Planck2016_XXIX}. Typical values for the emissivity index are in the range $1.2<\beta<2.2$ \citep{Planck2013_XI}.

\subsection{Aperture Photometry}
\label{sec:photometry}

Aperture photometry is a standard method of measuring the flux density of a source by integrating the brightness over the primary aperture enclosing the source and subtracting a background level measured from a background aperture. It has the advantage that it is not affected by the zero-level of the maps, although the primary and background apertures of the source must be spatially well defined in order to obtain accurate flux densities.

In order to map flux densities in \mbox{$\lambda$ Orionis}, aperture photometry with a common background aperture and a moving primary aperture is used. The background aperture is set as a fragmented annulus outside the ring, centred at G195.70$-$11.60 and with inner and outer radii of $7.\!^{\circ}0$ and $8.\!^{\circ}0$, respectively. The annulus is angularly limited to $150-\ang{200}$ and $300-\ang{325}$, where \ang{0} is Galactic North, as shown in Fig.\,\ref{fig:multifrequency}. This ensures that the background is measured in regions of minimal foreground contamination, avoiding the Orion region towards the west, dust emission towards Taurus on the east and emission from the Galactic plane towards the north.

The moving primary aperture is the flux of each \texttt{HEALPix} \citep{Gorski2005} pixel at $N_{\rm side}=64$ so that pixels are quasi-independent of each other. Maps at $N_{\rm side}=256$ are also produced, but these are only used for visualisation purposes in Section \ref{sec:parameter_maps} and $N_{\rm side}=64$ maps are used for all the analyses presented. The median background is subtracted from the primary aperture at each frequency. The flux density uncertainties are a combination of the root mean square of background fluctuations and the effective calibration errors shown in Table\,\ref{tab:surveys}. This is a reasonable approach due to the very high signal-to-noise ratio of the maps in the region at degree-scales, where the dominant uncertainty in flux density measurements comes from absolute calibration uncertainties. This results in parameter maps with a resolution given by the convolution of a $\ang{1}$ FWHM Gaussian and the pixel size, giving an effective resolution of FWHM$\,\approx \ang{1}$.

Colour corrections to account for the spectral effects of using finite bandwidths are applied iteratively using bandpass measurements for \mbox{C-BASS}, \mbox{QUIJOTE}, \textit{WMAP} and \textit{Planck}, and using top-hat bandpasses where bandpass measurements are unavailable. These corrections are at the level of less than a few percent, and under $1$ per cent in most cases. 

\subsection{MCMC SED Fitting}
\label{sec:mcmc_fitting_method}

SEDs are fitted using the \texttt{EMCEE} ensemble sampler \citep{Foreman-Mackey2013} backend, based on \mbox{\cite{Goodman2010}}, and using a standard maximum Gaussian likelihood function. The main advantage of using MCMC fitting over a least-squares method is that MCMC enables the full exploration of parameter space, which is necessary to assess correlations between variables, the stability of solutions (e.g., multimodal solutions, skewed distributions, etc.) and other factors that can bias least-squared fitting methods. This is especially important in the 7-dimensional parameter space fitted in this paper. Since the MCMC algorithm provides a joint-fit of all the relevant parameters, the marginalised final uncertainties take into account any correlations between parameters.

In order for the results to be derived entirely from the data, informative priors are not used. Instead, hard priors based on physical lower and upper limits are set and any quantity affected by the priors is masked from the resulting parameter maps. This is done by masking pixels where the parameter distribution fitted to the pixel is less than $3\,\sigma$ away from the edge of the prior range of the parameter, meaning that the remaining parameter distributions are not affected by the priors. Limits of $10<T_{\rm d}<100\,\si{\kelvin}$, $5<\nu_{\rm AME}<70\,\si{\giga\hertz}$ and $0.2<W_{\rm AME}<1.0$ are used, based on the typical physical constraints in \textit{Planck} thermal dust maps \mbox{\citep{Planck2013_XI}} and the statistical study of AME regions in \mbox{\cite{Planck2014_AME}}. No priors are set in any other parameter, including component amplitude parameters $A_{\rm AME}$, $\mathrm{EM}$ and $\tau_{\rm 353}$. This ensures that detections of each amplitude are driven purely by the data and not biased by positivity priors. The limits on $W_{\rm AME}$ parameter ensure that the shape of the spectrum is within the distribution of widths found for theoretical \textsc{SpDust2} templates, which are in the range $\sim 0.4 < W_{\rm AME} < 0.7$. This suggests that any log-normal widths above 1.0 or below 0.2 are unphysical and can thus be masked. The AME width is particularly hard to constrain in the centre of the ring, where only a marginal detection of the amplitude of AME is made.

\begin{figure*}
	\begin{center}
    \setbox1=\hbox{\includegraphics[height=20.7cm]{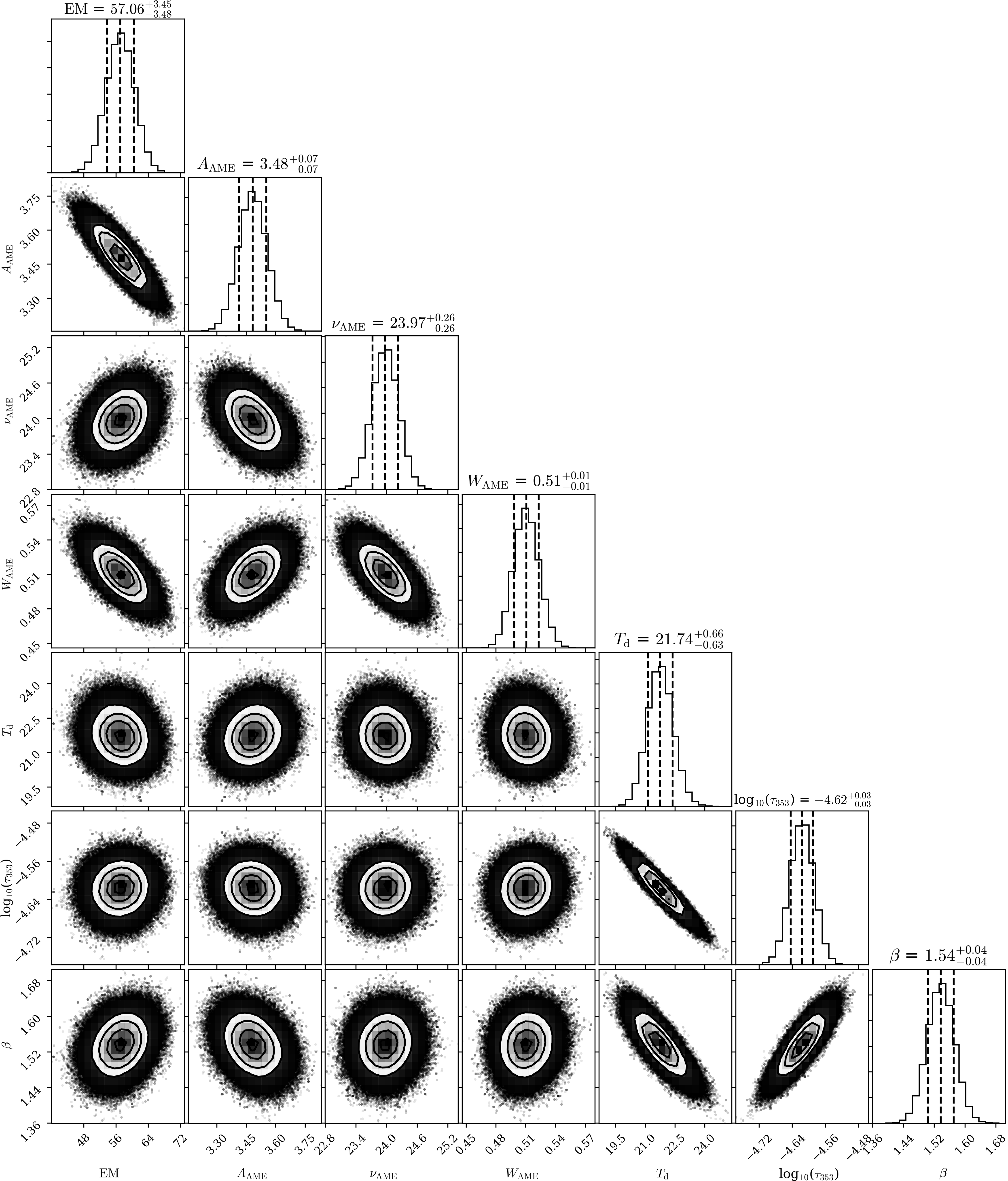}}
    \includegraphics[height=20.7cm]{corner.png}\llap{\raisebox{12.7cm}{\includegraphics[height=8.0cm]{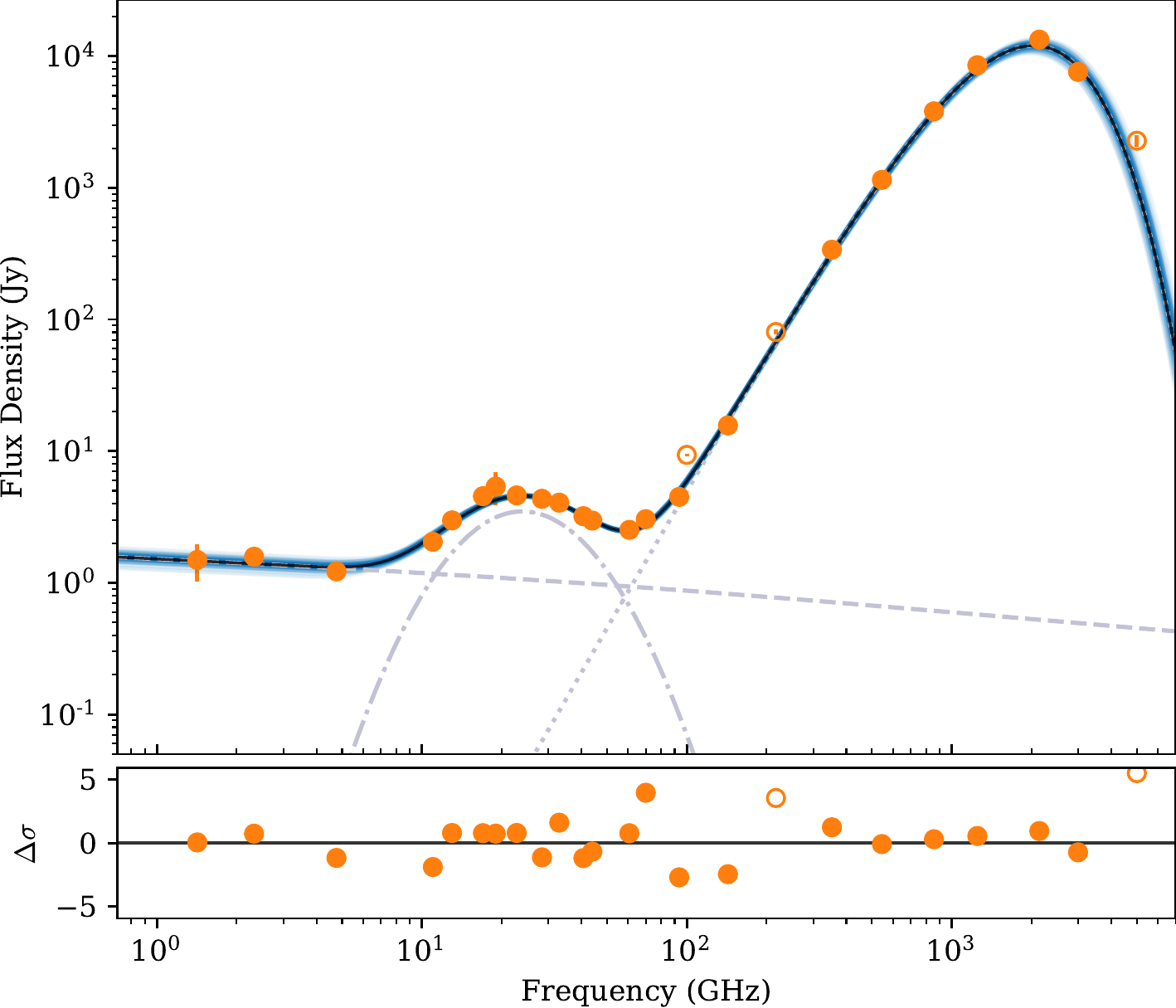}}} \\
	\caption{Corner plot (\textit{left}) and SED (\textit{top-right}) for G194.77$-$15.71 near dark cloud B223 (region \textit{B}), which is one of the brightest AME cores. The corner plot shows 2D histograms of the 1.5 million walker positions after the burn-in period for every parameter pair, with a total position histogram at the end of each row, showing the mean and $1\,\sigma$ bounds of each parameter. The SED (\textit{top-right}) shows flux densities in \textit{orange} and the best fit curve (\textit{blue}) to a free-free (\textit{dash-dash}), AME (\textit{dash-dot}) and thermal dust (\textit{dot-dot}) models. The fit residuals are shown below the SED as $\Delta\sigma$, which indicates the number of standard deviations each data point is away from the best fit model. Note that most photometric uncertainties are smaller than the data points, and that unfilled circles are excluded from the fit.}
	\label{fig:mcmc}
\end{center}
\end{figure*}

In total, 300 walkers are used, each walking a total of 10000 steps, of which the first 5000 are removed as ``burn-in'' steps. All fits converge in $< 1000$ steps, with most reaching a steady state after $\sim200$ steps, justifying the choice of using 5000 ``burn-in'' steps. This leaves a total of 1.5 million samples in the posterior distribution, which are shown in Fig.\,\ref{fig:mcmc} for a SED of dark cloud B223 in region \textit{B}, which is one of the brightest AME cores. The number of steps to convergence is also reduced by initializing initial walker positions with a least squares fit and then randomizing the resulting parameters with a $50$ per cent Gaussian distribution. Walkers stuck outside of priors are also removed from the posterior distribution. Step acceptances between $40\,$ and $50$ per cent are observed in the parameter maps, indicating an appropriate step size. The correlation plots in Fig.\,\ref{fig:mcmc} show that AME and thermal dust parameter MCMC distributions are very weakly correlated, while the strongest correlations are observed among the three thermal dust parameters. Linear correlations can also be seen between $A_{\rm AME}$ and $\mathrm{EM}$, as well as between the three AME parameters. These are the correlations that could bias a simple least-squares fit.

\subsection{Validation Tests}
\label{sec:validation}

A number of simulations were run to test the robustness and the ability of the MCMC pipeline to recover true sky parameters. The intrinsic decrease of the peak frequency gradient in the region was also tested, ruling out the possibility that degeneracies between $\nu_{\rm AME}$ and other parameters contribute to the peak frequency gradient.

The basis of the simulations is the generation of SEDs from sets of physical parameters, which are then refitted using the MCMC pipeline. The simulated SEDs are created by calculating the flux densities and randomizing them by the effective calibration error shown in Table\,\ref{tab:surveys}. The uncertainties on the simulated SED flux densities are set by the calibration uncertainty on each parameter map, since this is the main source of uncertainty in the analysis.

Different input sets of physical parameters were tested. Firstly, the resulting parameter maps obtained were used as the input of the simulations to test the stability of the results. The recovered parameter maps were consistent with our results in Fig.\,\ref{fig:parameter_maps} for each parameter, showing that the pipeline is robust since it can reproduce our results. An example of measured vs. simulated parameters is shown in Fig.\,\ref{fig:peak_frequency} for the measured (\textit{orange} markers in the \textit{bottom} plot) and simulated (\textit{grey} markers) AME peak frequencies, showing that the intrinsic decrease of the peak frequency with radius was recovered. More importantly, in order to show that the fits are not biased and confirm that the fitting pipeline recovers the true sky parameters, test input scenarios were used. In particular, we tested a scenario where every parameter is changing but the peak frequency is forced to have a constant value of $\nu_{\rm AME}=21\,\si{\giga\hertz}$. The simulation recovered the fixed peak frequency despite variations in other parameters, implying that the peak frequency gradient is not a consequence of the degeneracy between the radially decreasing free-free level. General cases of combinations of physical parameters in the region were also tested in order to check that the MCMC fitting routine can recover the true sky parameters. One such case was a low signal-to-noise AME scenario, where the input values are consistent with those recovered by the pipeline.


\section{Results and Discussion}
\label{sec:discussion}

\subsection{Physical Parameter Maps}
\label{sec:parameter_maps}

\begin{figure*}
	\begin{center}
		\includegraphics[width=\textwidth,angle=0]{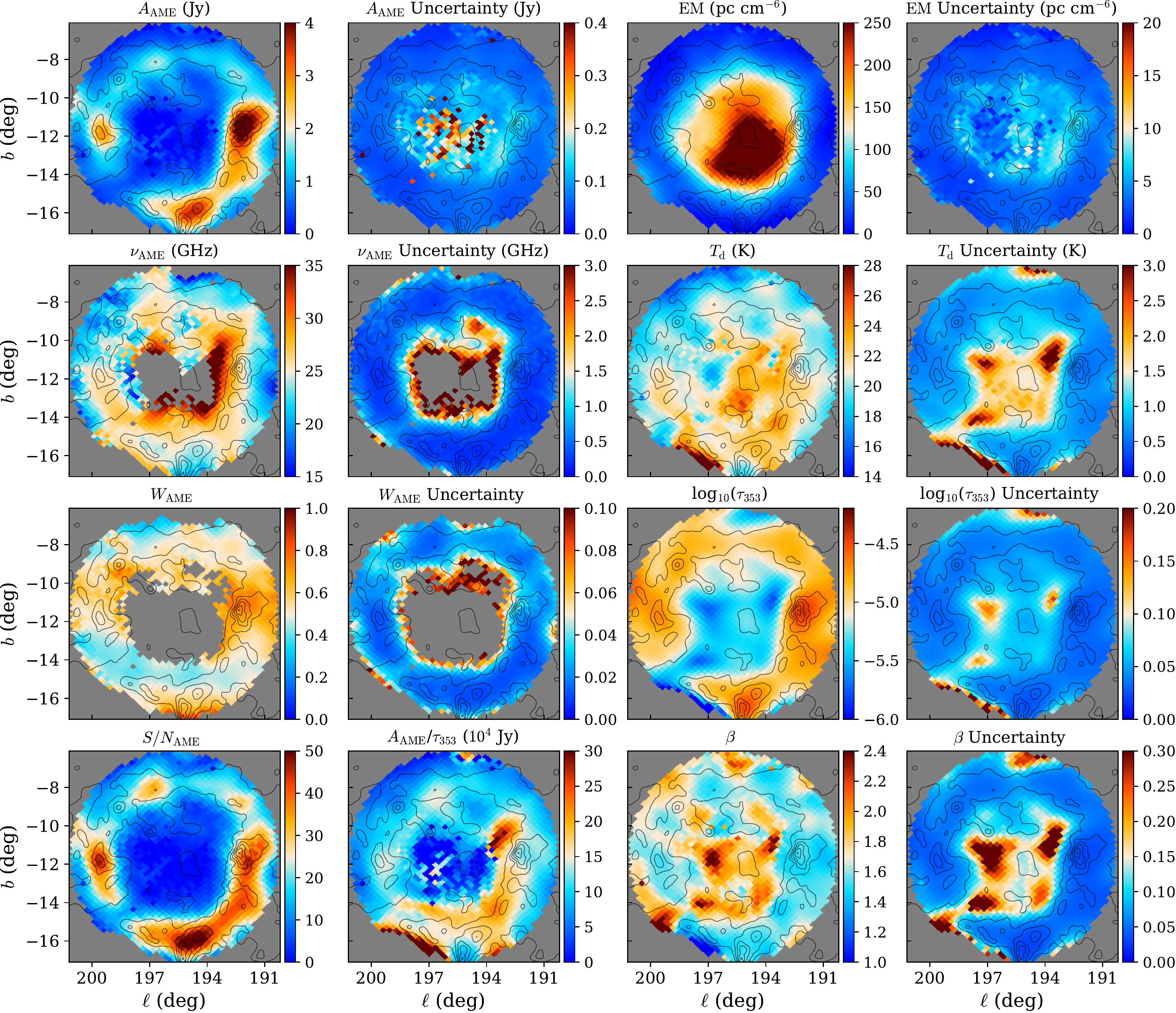}
		\caption{Best-fit \mbox{$\lambda$ Orionis} parameter maps in Galactic coordinates centred on G195.7$-$11.6, spanning $12.\!^{\circ}3\times12.\!^{\circ}3$. Regions a beam-width away from the background apertures are masked, as well as any point where parameters are in any way affected by fitting priors. The signal-to-noise ratio of AME is denoted by ${S/N}_{\rm AME}$, where the detection of AME is in excess of $20\,\sigma$ around the entirety of the ring. The contours are the \textit{Planck} $857\,\si{\giga\hertz}$ map smoothed to a FWHM of $20\,$arcmin with 6 evenly-spaced levels between 14.7 and 58.9$\,$MJy$\,\si{\steradian}^{-1}$ (the maximum flux density in the region) in steps of 8.84$\,$MJy$\,\si{\steradian}^{-1}$. The $20\,$arcmin FWHM is chosen in order to preserve sub-degree details while making contours easier to visualize.}
		\label{fig:parameter_maps}
	\end{center}
\end{figure*}

The physical parameter maps extracted from the MCMC SED fits are shown in Fig.\,\ref{fig:parameter_maps}, together with their uncertainties. An AME signal-to-noise map is also presented as ${S/N}_{\rm AME}$, which shows that AME is detected with a statistical significance between $\approx20\,\sigma$ and $\approx50\,\sigma$ around the ring, and typical ${S/N}_{\rm AME}$ between $\approx3\,\sigma$ and $\approx7\,\sigma$ in the H\textsc{ii} bubble due to the lower amplitude of AME in the region. All parameters with detections of each single parameter under $3\,\sigma$ are masked in Fig.\,\ref{fig:parameter_maps}.

The overall environment of the region is reflected by the thermal dust and free-free emission parameters. The emission measures recovered trace the radially decreasing ionization from \mbox{$\lambda$ Orionis}, and they also reflect the clear distinction between the inner H\textsc{ii} region and the dust ring at a radius of $\approx4^{\circ}$. The strongest emission measures are skewed towards the south-east, with a steep reduction in free-free emission between regions \textit{A} and \textit{B} corresponding to the photodissociation region between the ring and the H\textsc{ii} shell. The effect of the radiation field from \mbox{$\lambda$ Orionis} is also imprinted in the dust temperature map, with $T_{\rm d}$ varying from $\approx27\,\si{\kelvin}$ near the central star to $\approx18\,\si{\kelvin}$ in the ring in a less uniform radially outward fashion. The optical depth map, $\tau_{353}$, shows a very high degree of correlation with the thermal dust emission, as expected. The emissivity index, $\beta$, is also found in the range $1.2 < \beta < 2.2$, and in the range $1.4 < \beta < 1.8$ for most of the ring, consistent with typical measurements of molecular clouds in \cite{Planck2013_XI}.

A strong correlation can be seen between the AME amplitude and the \textit{Planck} $857\,\si{\giga\hertz}$ map denoted by the contours overlaid on each parameter map, with the three cores labelled regions \textit{A}, \textit{B} and \textit{C} in Fig.\,\ref{fig:multifrequency} all having significant AME. The correlation between thermal dust and AME is a well-known relation since AME from spinning dust, if assumed to be the dominant mechanism, originates from grains that are typically mixed with the larger grains that are dominant in thermal dust emission.

The peak frequency of AME is also seen to be higher towards the centre of the ring, going from $\approx35\,\si{\giga\hertz}$ in the H\textsc{ii} bubble to $\approx21\,\si{\giga\hertz}$ in the ring, with typical peak frequencies of $\approx24\,\si{\giga\hertz}$ and dust temperatures of $\approx21\,\si{\kelvin}$ at the three main molecular cloud cores. A significant radially outward decrease of the peak frequency can be seen in the parameter maps. This is discussed in detail in Section\,\ref{sec:ame_variations}. Around the ring, the width parameter is seen to physically vary in a significant way between $0.4 < W_{\rm AME}< 0.7$, with wider distributions at the cores of regions \textit{A} and \textit{B}. While this parameter is potentially a proxy for the variety populations of spinning dust grains due to the predicted widening of the AME curve from the superposition of different populations, no strong correlation between it and the individual thermal dust emission parameters is observed. No correlations between $W_{\rm AME}$ and the AME peak frequency or amplitude are observed either.

The $\tau_{\rm 353}$- normalized AME amplitudes trace the per-grain emission of AME, being a proxy for $j_{\nu}/n_{\rm H}$ where $j_{\nu}$ is the emissivity, and show higher per-grain emission by a factor of a few near the edges of the PDR enclosed by the ring.

It is worth noting that intrinsic spatial correlations exist between parameters due to the geometry. An example of this is free-free emission and the dust optical depth, which are inversely correlated. Correlations between pairs of parameters and their interpretations are discussed in Section\,\ref{sec:spinning_dust}.

\subsection{{\textsc{Commander}} Comparison}
\label{sec:commander}

\begin{figure}
	\begin{center}
		\includegraphics[width=0.235\textwidth,angle=0]{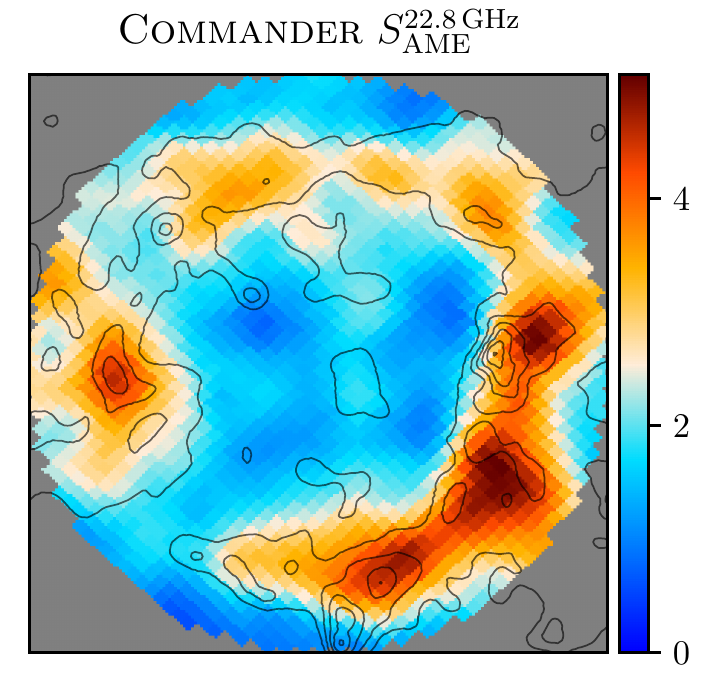}
		\includegraphics[width=0.235\textwidth,angle=0]{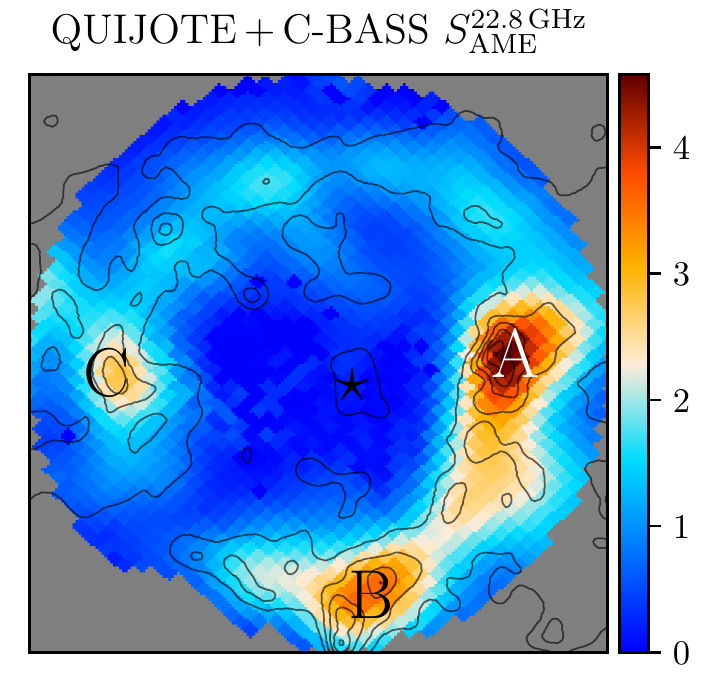}
		\includegraphics[width=0.235\textwidth,angle=0]{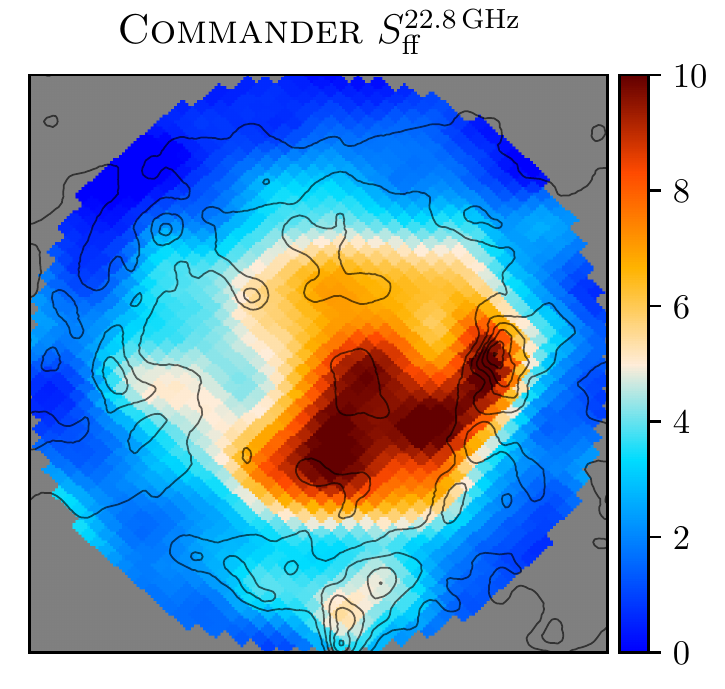}
		\includegraphics[width=0.235\textwidth,angle=0]{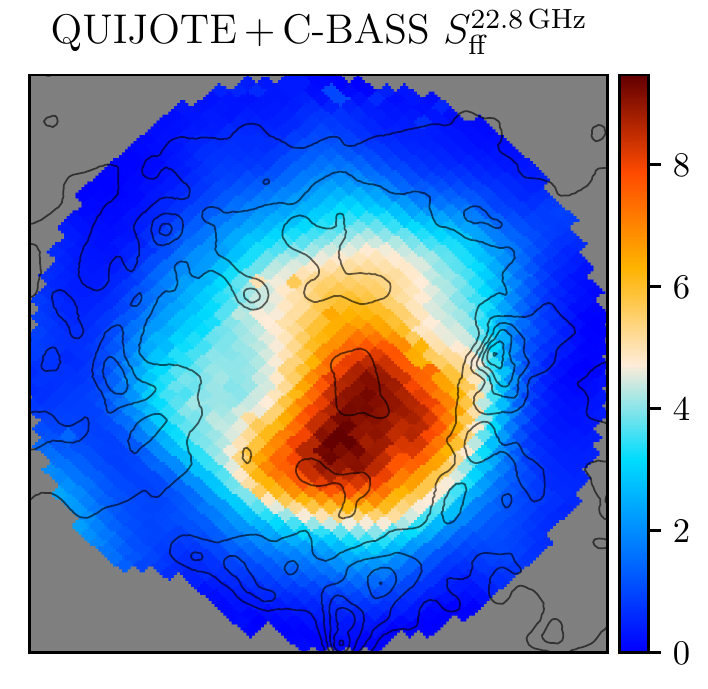}
		\caption{Comparison between the \protect\cite{Planck2015_X} \textsc{Commander} maps and the MCMC-SED-separated maps using \mbox{QUIJOTE} and \mbox{C-BASS} for AME (\textit{top} plots) and free-free emission (\textit{bottom} plots). All flux density scales correspond to Jy, evaluated at $22.8\,\si{\giga\hertz}$ for an aperture with a FWHM of $\ang{1}$. The contour levels are the same as those shown in Fig.\,\ref{fig:parameter_maps}. A reminder of the location of key regions A, B and C is shown in the top-right figure, where the position of O8 III star \mbox{$\lambda$ Orionis} is indicated by the star-shaped marker.}
		\label{fig:commander}
	\end{center}
\end{figure}

\textsc{Commander} \citep{Eriksen2008b} is a widely used algorithm for joint component separation and CMB power spectrum estimation based on Gibbs sampling, which uses parametric models of the foreground signals to separate them. Here, we use the \cite{Planck2015_X} implementation of the \textsc{Commander} foreground separation maps available in the ESA \textit{Planck} Legacy Archive\,\footnote{\url{http://pla.esac.esa.int/pla/}} to assess their reliability. In this section, our focus is to show that component separation methods in general suffer from degeneracies. We focus on \textsc{Commander}, this being one of the most used separations for foreground studies since it is one of the few that provide individual component low frequency maps. In particular, we are interested in assessing the improvements in the separation of AME and free-free with the better low-frequency coverage provided by \mbox{C-BASS} and \mbox{QUIJOTE}. The \textit{Planck} implementation uses two \textsc{SpDust2} diffuse cold neutral medium templates to model the AME: one with a varying peak frequency, and another with a constant peak frequency of $33.35\,\si{\giga\hertz}$ to account for broadening due to a superposition of components. Note that the results of this assessment are independent of the method chosen for CMB subtraction (\textsc{NILC}) since the method chosen does not affect the results in any significant way.

This study serves as a first demonstration of the improved intensity component separation with the addition of \mbox{C-BASS} and \mbox{QUIJOTE}. We do this by comparing the expected free-free and AME levels derived from our parameter maps in Fig.\,\ref{fig:parameter_maps} with the \cite{Planck2015_X} \textsc{Commander} estimates, which only use data from \textit{Planck}, \textit{WMAP} and the $0.408\,\si{\giga\hertz}$ survey by \cite{Haslam1982}. A direct comparison in the region is shown in Fig.\,\ref{fig:commander}. While the arbitrary zero-levels of the two maps are different due to the way in which they are processed, the overall shapes of AME and free-free emission are similar. However, there are evident differences in the shape of the inner free-free region and in the relative brightness between the two main dark cloud clusters in regions \textit{A} and \textit{B}, as well as the appearance of a strong AME region in the \cite{Planck2015_X} \textsc{Commander} map joining the two main cores. Individual examples point to clear leakages of AME into free-free emission and vice versa, which can be seen by eye. However, leakages primarily occur from AME into free-free in this region, which is consistent with \mbox{QUIJOTE} studies of the Taurus region by \cite{Poidevin2019} and observations of W43, W44 and W47 by \cite{Genova-Santos2017}. These leakages originate from degeneracies between components due to the limited multi-frequency data available to the \cite{Planck2015_X} \textsc{Commander} separation, and they were observed and discussed in \cite{Planck2015_XXV, Planck2018_IV}.

A clear overestimation of AME is seen in the \textsc{Commander} maps between regions \textit{A} and \textit{B} as a distinct AME feature. This feature is not expected since it does not correspond to any bright regions of thermal dust emission. An example of leakage of AME into free-free can be seen in region \textit{B} in the \textsc{Commander} free-free map, which results in a bright spot of free-free emission strongly correlated with thermal dust in an environment where it is not physically expected. A similar case of leakage of AME into free-free is observed to a greater extent in region \textit{A}, where the majority of AME in the centre of region \textit{A} is lost to free-free emission. This shifts the centre of AME in \textsc{Commander} in region \textit{A} away from the maximum thermal dust brightness, represented by the contours. This results in the AME bright spot in region \textit{A} not appearing as a distinct, dominant feature in the \textsc{Commander} map, whereas in our map AME in region \textit{A} correlates strongly with thermal dust emission, as reflected by the contours. At the same time, this case of leakage in region \textit{A} increases free-free emission significantly, giving the overall free-free emission a tilted \textit{N}-shape in the \textsc{Commander} map, which is not observed in our free-free map. Since ionization in the region is dominated by a single central star (marked in Fig. \ref{fig:multifrequency}), a smooth radial decrease in the free-free level like the one in our map is expected over the structure in the \textsc{Commander} free-free map.

Overall, a key physical argument for our separation being more reliable are the stronger correlations between AME and thermal dust emission in our maps. The Spearman correlation coefficients between thermal dust emission at $857\,\si{\giga\hertz}$ and our AME map, evaluated at $N_{\rm side}=64$, is $0.89$. With the \cite{Planck2015_X} \textsc{Commander} map, it drops to $0.85$. Similarly, the Spearman correlation coefficient between the total fluxes at $22.8\,\si{\giga\hertz}$ as measured in \textsc{Commander} and our analysis including \mbox{C-BASS} and \mbox{QUIJOTE} is $0.98$, meaning that the total fluxes are very highly correlated. However, mixing between free-free and AME is reflected in the lower coefficients between the two AME maps, $r_s=0.81$, and the two free-free maps, $r_s=0.89$, where $r_s$ denotes a Spearman correlation coefficient. This highlights the importance of breaking degeneracies through low-frequency data such as \mbox{C-BASS} and \mbox{QUIJOTE}, and in particular the combination of several experiments, where the overall constraints in component separation are greater than their individual contributions. In this case, \mbox{QUIJOTE} and \mbox{C-BASS} work together, with \mbox{QUIJOTE} primarily constraining the peak frequencies and amplitudes of AME and \mbox{C-BASS} constraining the free-free level, which in turn contributes to the measurement of $A_{\rm AME}$. This case study of improved low-frequency component separation also emphasizes the limitations of studies that make use of the \mbox{\cite{Planck2015_X}} implementation of the \textsc{Commander} maps, which do not include additional low-frequency data. This is particularly important in this region, where the two highest signal-to-noise AME features are very different from the \textsc{Commander} estimates.

\subsection{AME Peak Frequency Variations}
\label{sec:ame_variations}

\begin{figure}
	\begin{center}
		\includegraphics[width=0.47\textwidth,angle=0]{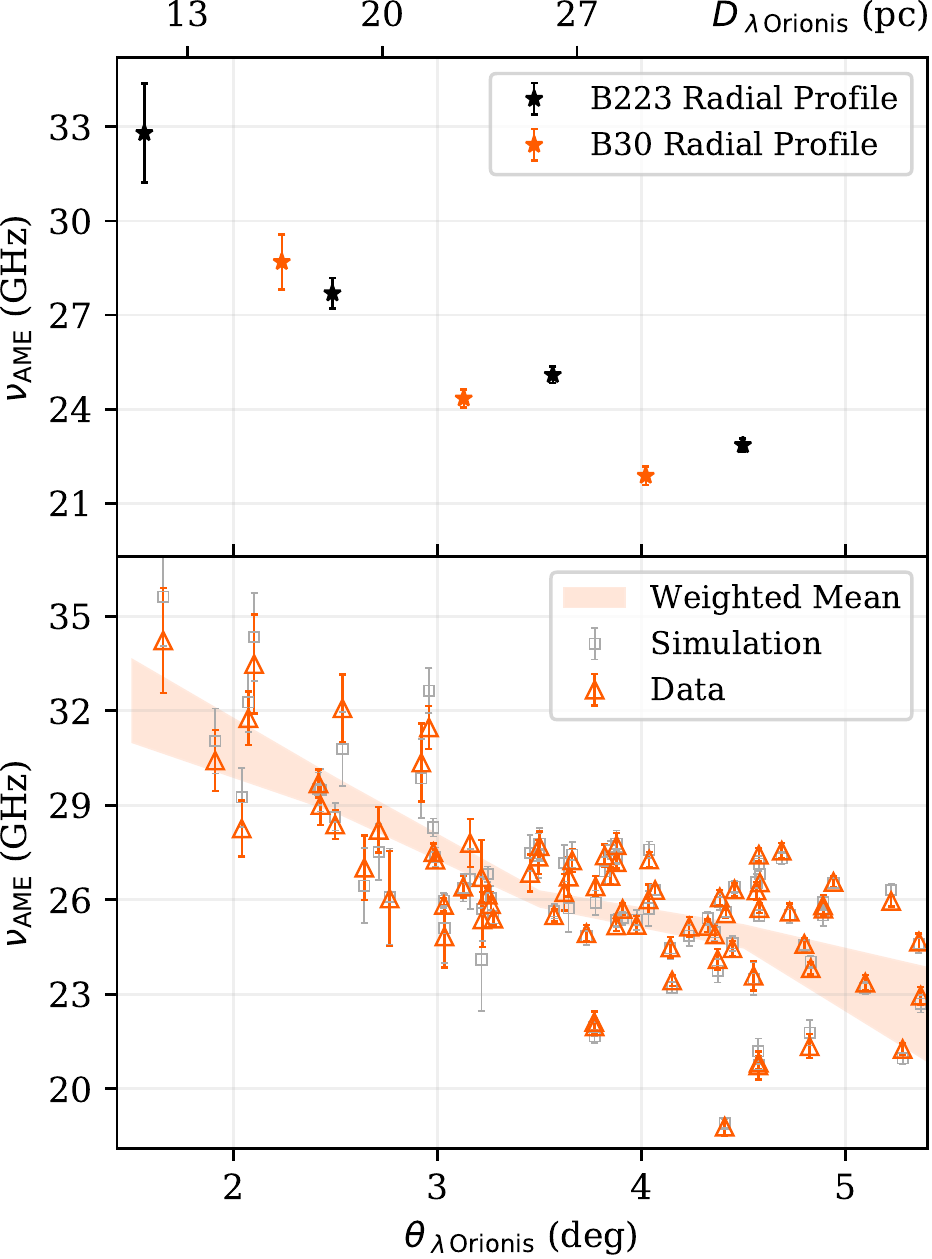}
		\caption{AME peak frequency as a function of angular distance from star \mbox{$\lambda$ Orionis}. Data points are drawn from maps at \mbox{$N_{\rm side}=64$}, making each point quasi-independent since the pixel solid angle is $\approx0.84\,\si{\deg}^2$. The \textit{x}-axis is also shown at the top as a function of physical distance from \mbox{$\lambda$ Orionis} based on \protect\cite{Schlafly_2014}, accurate to $\approx10$ per cent. \textit{Top}: two radial peak frequency profiles spaced by one degree emanating from star \mbox{$\lambda$ Orionis}, with the black radial profile crossing molecular cloud B223 near the centre of region \textit{B} and the orange radial profile crossing molecular cloud B30 in region \textit{A}, which are the areas with the highest signal-to-noise ratio. \textit{Bottom}: given the approximately circular shape of the source, we bin the source radially, showing the overall reduction in peak frequency in the region in all directions in \textit{orange}, along with the simulated peak frequencies from the validation tests detailed in Section\,\ref{sec:validation} shown in \textit{light grey}, which show that the peak frequency decrease is recovered. The weighted mean and the $1\,\sigma$ scatter are shown in $1^{\circ}$ bins as a shaded \textit{orange} region.}		\label{fig:peak_frequency}
	\end{center}
\end{figure}

We report the first observational evidence of spatial variations in the AME peak frequency in a single region. Fig.\,\ref{fig:peak_frequency} shows $\nu_{\rm AME}$ as a function of the angular distance from star \mbox{$\lambda$ Orionis} along radii running across two specific molecular clouds (\textit{top}) and the overall trend in the entirety of the region (\textit{bottom}, \textit{orange} markers). Points with separations less than $\approx4^{\circ}$ sit inside the ring, in the inner regions on the H\textsc{ii} bubble. A clear overall decrease of the peak frequency is observed the further we go from \mbox{$\lambda$ Orionis}, with each radial profile (\textit{top}) having a more well-defined functional form than the overall population (\textit{bottom}), with a hint of non-linearity in the spatial decrease of $\nu_{\rm AME}$. It is also clear that the angular distance is not the only parameter that determines the decrease in $\nu_{\rm AME}$, since the radial profile traversing B30 (region \textit{A}) is steeper than the B223 (\textit{B}) profile. In region \textit{C}, this pattern is not visible, in part due to the larger peak frequency uncertainties in the region and likely due to environmental differences in the ionization reflected by the free-free map in Fig.\,\ref{fig:commander}. For this reason, the fact that the overall shape of the population (\textit{bottom} plot) shows a higher scatter than each one of the radial profiles it is to be expected, since the decrease in $\nu_{\rm AME}$ would depend on local environmental parameters. Such environmental factors may include the intrinsic difference in the chemistry and sizes of grains in different molecular clouds, regions of active star formation with their own radiation fields around \mbox{$\lambda$ Orionis} \mbox{\citep{Dolan2002}} where local interstellar radiation field (ISRF) contributions are comparable to those from \mbox{$\lambda$ Orionis} itself, and the fact that certain lines of sight intersect with foregrounds structures in the 3D shell around \mbox{$\lambda$ Orionis}, giving a beam-integrated measurement of their properties. In addition, the region is only approximately circularly symmetric.

While there are well-known differences in peak frequencies between known AME sources in the sky, such as between Perseus with $\nu_{\rm AME}=27.8\pm0.3\,\si{\giga\hertz}$ and the California nebula with $\nu_{\rm AME}=50\pm17\,\si{\giga\hertz}$ \citep{Planck2014_AME}, detecting changes in the peak frequency in a single region enables us to test spinning dust models by linking local environmental parameters to AME in the region and assessing the fundamental relations between them. With $\approx\ang{1}$ resolution, studies like \mbox{\cite{Planck2014_AME}} that look at smaller sources do not generally have enough resolution to resolve spatial variations, and beam dilution can affect the derived AME properties. However, spatial variations in AME amplitude are expected and have been observed in arcminute-scale studies, such as those of Lynds nebulae \citep{Casassus2006, Scaife2009}, $\rho$ Ophiuchi \citep{Casassus2008, Arce2019}, H\textsc{ii} region RCW175 \mbox{\citep{Dickinson2009a, Battistelli2015}}, Orion \citep{Dickinson2010} and Perseus \citep{Tibbs2013}. By observing spatial variations in the amplitude of AME, these studies suggest that spatial variations in all properties of AME are widespread at sub-degree scales, including spectral variations, which can be linked to local environmental parameters. Recently, \cite{Casassus2020} have reported spectral variations of the AME peak frequency across the $\rho$ Ophiuchi photodissociation region, with higher peak frequencies found closer to the ionizing star. This is another hint that AME spectral variations are common, and that PDRs are good environments to search for them.

The relations between the peak frequency and environmental parameters in \mbox{$\lambda$ Orionis} and their interpretation are discussed in Section\,\ref{sec:spinning_dust}.

\subsection{Spinning Dust}
\label{sec:spinning_dust}

\begin{figure*}
	\begin{center}
		\includegraphics[width=\textwidth,angle=0]{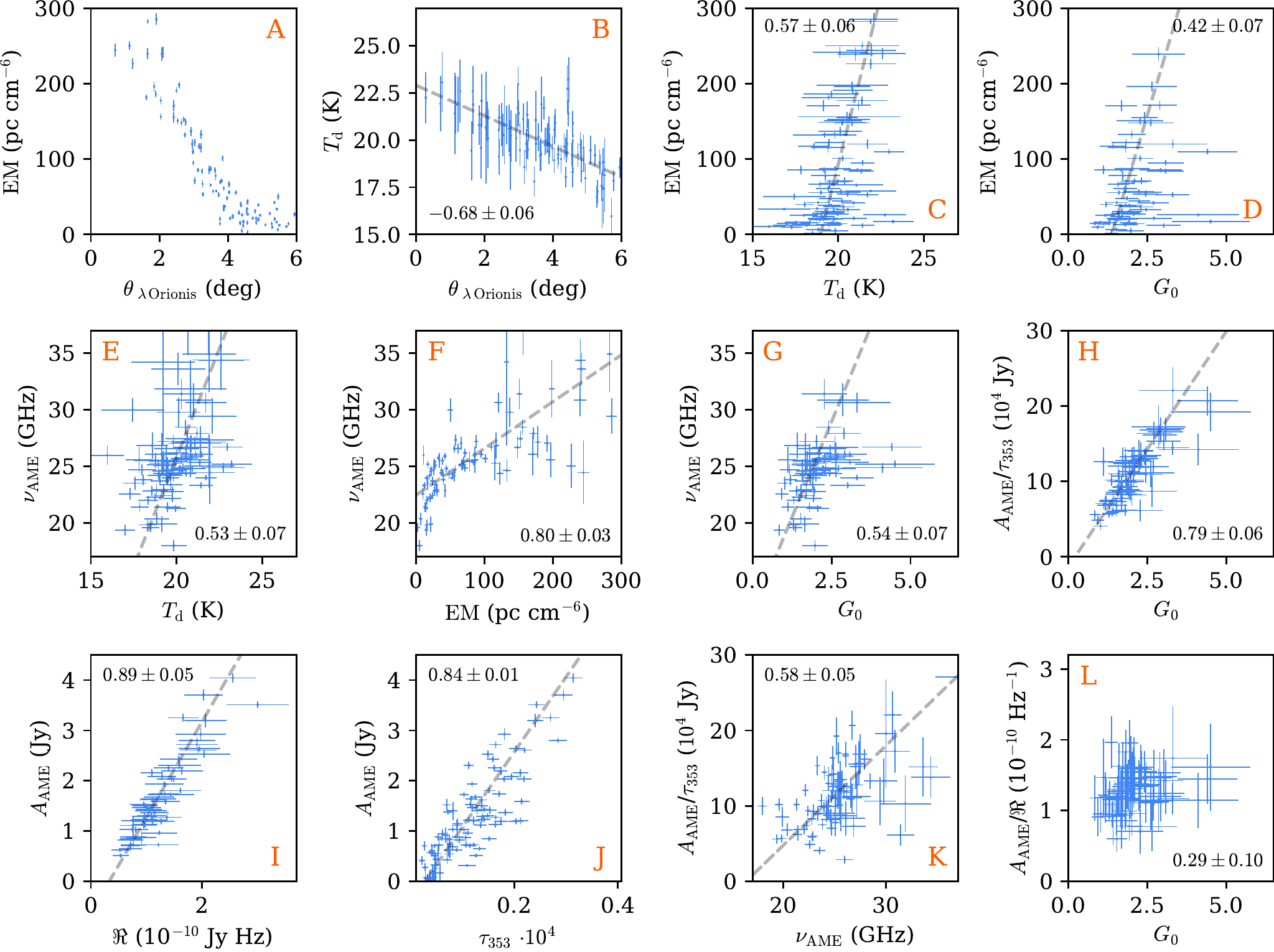}
		\caption{Correlation plots between the pairs of variables shown in Fig.\,\ref{fig:parameter_maps} with a significance greater than $3\,\sigma$. Each data point is from an \mbox{$N_{\rm side}=64$} pixel. The angular distance from the position of OB star \mbox{$\lambda$ Orionis}, located at G195.05$-$12.00, is denoted by $\theta_\mathrm{\lambda\,Orionis}$. The numbers printed denote the Spearman correlation coefficients for each pair of parameters. Dashed lines represent linear models fitted to data with significant ($>5\,\sigma$ from zero) Spearman correlation coefficients.}
		\label{fig:correlations}
	\end{center}
\end{figure*}

We now compare the various physical parameters in Fig.\,\ref{fig:parameter_maps} to each other in order to assess the fundamental relations between them in the region and relate them to spinning dust. In particular, the relation between AME parameters and environmental parameters such as the relative strength of the interstellar radiation field and thermal dust emission are explored. A selection of plots between parameters is shown in Fig.\,\ref{fig:correlations}. The ISRF strength is estimated using the proxy $G_0=\left ( T_{\rm BG}/17.5\,\si{\kelvin} \right )^{4+\beta_{\rm BG}}$ \mbox{\citep{Mathis1983}}, where $T_{\rm BG}$ is the temperature of large grains relative to the average value of $17.5\,\si{\kelvin}$ and $\beta_{\rm BG}$ is the emissivity index of large grains. We approximate these parameters using the values obtained from SED fitting, i.e., $T_{\rm BG} \approx T_{\rm d}$ and $\beta_{\rm BG} \approx \beta$. We also calculate the thermal dust radiance, $\Re=\int_{0}^{\infty}S_{\rm d}(\nu)\,\rm{d}\nu$, and its uncertainties by directly integrating the fitted thermal dust curve.

The first two plots in Fig.\,\ref{fig:correlations}, labelled \textit{A} and \textit{B}, show the radial dependence of the emission measure and dust temperature in the region. A sharp transition to a flat free-free level can be seen beyond $\approx4^{\circ}$, corresponding to the position of the PDR, while the dust temperature decreases with angular distance from the central ionizing star, as expected. These two variables are therefore correlated as seen in plot \textit{C} through the effects imprinted by the radially decreasing radiation field produced by star \mbox{$\lambda$ Orionis}, with a Spearman coefficient of $0.57\pm0.06$ as shown in plot \textit{C}. An example of the correlation between $\mathrm{EM}$ and $G_0$ is shown in plot \textit{D}. Both linear models and power laws (i.e., $y = A\,x^{\alpha}+C$) are fitted to the data in order to determine the non-linearity of the relations. This is done using the \texttt{Scipy} orthogonal distance regression package based on \cite{Boggs1990}, which considers uncertainties in both axes. In all cases except for plot \textit{A}, there is little evidence for deviations from $\alpha = 1$, so linear models are fitted and shown instead. In the case of plot \textit{A}, non-linearity is expected due to the discontinuity between the ionized and non-ionized environments separated by the PDR, which is why a linear fit is not shown.

Plot \textit{E} shows a correlation between $\nu_{\rm AME}$ and $T_{\rm d}$, with a Spearman correlation coefficient of $0.53\pm0.07$ and a linear gradient of $3.5\pm0.5\,\si{\giga\hertz}\,\si{\kelvin}^{-1}$. No significant deviations from linearity are found by fitting a power law. This correlation is perhaps surprising initially, since \textsc{SpDust} predictions, shown in \mbox{\cite{Ali-Hamoud2009}}, do not predict a direct dependence of $\nu_{\rm AME}$ on $T_{\rm d}$ in the range of radiation field strengths we observe in \mbox{$\lambda$ Orionis}. In fact, \cite{Ali-Hamoud2009} show that the rise of peak frequency with radiation field occurs at $G_{0}\gtrsim 100$, while we expect $1\lesssim G_{0}\lesssim 10$ in the region at degree scales. However, a range of environmental factors can lead to making this trend a reality, such as the effect of the radiation field driving the dust grain size distributions and dust temperatures in the region. 

Two similar correlations can be seen in plot \textit{F} between $\nu_{\rm AME}$ and $\mathrm{EM}$, and in plot \textit{G} between $\nu_{\rm AME}$ and $G_0$. These have Spearman coefficients of $0.80\pm0.03$ and $0.54\pm0.07$, respectively. While the fact that all four parameters $\nu_{\rm AME}$, $\mathrm{EM}$, $T_{\rm d}$ and $G_0$ are correlated with each other makes it difficult to determine the cause of the correlation, the radiation field is a likely interpretation of this effect. The effects of the ISRF would be imprinted in all $\mathrm{EM}$, $T_{\rm d}$ and $G_0$, suggesting that the peak frequency changes as a result of the changes in the properties of the grains themselves as a consequence of the ISRF, such as dissociation into smaller grains and a greater energy budget for achieving higher rotational frequencies. These are the effects that may be dominating the correlation, assuming \textsc{SpDust} is a good model of spinning dust. The fact that correlation coefficient with $G_0$ is lower than with $\mathrm{EM}$ and similar to the coefficient with $T_{\rm d}$ is to be expected since $G_0$ is merely a proxy for the true local interstellar radiation field strength. In \cite{Planck2014_AME}, measured values for $\nu_{\rm AME}$ are generally similar and smaller than those found inside the H\textsc{ii} region in \mbox{$\lambda$ Orionis}, with the most prominent outlier being the California nebula with $\nu_{\rm AME}=50\pm17\,\si{\giga\hertz}$. The California nebula is also associated with an H\textsc{ii} region where the radiation field is dominated by O7 star $\xi$ Persei \citep{Lada2009}. This is a hint that the radiation field is common and responsible, either directly or more likely indirectly through its effect on dust properties such as grain size distributions.

The next result is a correlation between the AME emissivity, approximated as $A_{\rm AME}/\tau_{353}$, and $G_0$, shown in plot \textit{H}, which also supports the idea that the ISRF is changing the properties of the grains by increasing ionization in spinning dust carriers, for example. This correlation was first reported in \cite{Tibbs2011,Tibbs2012}, and corroborated by \mbox{\cite{Planck2014_AME}} and Poidevin et al. (in prep.). The idea that the ISRF is changing grain properties is also supported by observations of LDN1780 by \cite{Vidal2019}, who concluded that a change in the grain size distribution can in turn also impact the AME emissivity.

Here, a correlation coefficient of $0.79\pm0.06$ is found between the two variables. This relation is not expected in \textsc{SpDust} models at the ISRF strengths observed \mbox{\citep{Ali-Hamoud2009}}, since the models predict that the emissivity changes very slowly over a large range of $G_0$ values, only rising above $G_0\gtrsim100$ in the case where the radiation field does not change the properties of the grains themselves. This implies that these properties are changing in reality. A similar relation is seen in plot \textit{K}, this time as a function of $\nu_{\rm AME}$, with a lower Spearman correlation coefficient of $0.58\pm0.05$. This is also expected in the region as a consequence of the correlation between $G_0$ and $\nu_{\rm AME}$ shown in plot \textit{G}.

The relations between $A_{\rm AME}$, the thermal dust radiance $\Re$ and $\tau_{353}$ are shown in plots \textit{I} and \textit{J}, which have Spearman coefficients of $0.89\pm0.05$ and $0.84\pm0.01$, respectively. $\Re$ shows a smaller scatter and higher correlation coefficient, making it the single best predictor of $A_{\rm AME}$ in the region. The AME radiance, not plotted, which is calculated from the integral of our fitted log-normal model, also correlates with $\Re$ with a Spearman coefficient of $0.82\pm0.06$, although a larger scatter is seen and expected due to AME modelling uncertainties. The coupling of radiances, and more clearly, the coupling between the AME amplitude and $\Re$ is consistent with the two emission mechanisms sharing the same power source (i.e., \mbox{$\lambda$ Orionis}), since their energy budgets correlate with one another. Radiance as the best $A_{\rm AME}$ predictor in the region is consistent with the analysis in \cite{Hensley2016}, with the caveat that their analysis uses the \citep{Planck2015_X} implementation of the \textsc{Commander} maps. In fact, in the region, \cite{Planck2015_XXV} find that the best correlations with the \textsc{Commander} AME estimates are with $\tau_{353}$ and the $545\,\si{\giga\hertz}$ map, not $\Re$. However, in other studies that use template fitting at higher latitudes, $\tau_{353}$ is found to be a better template than $\Re$ \citep{Dickinson2018AMEReview}. In observations of $\rho$ Oph by \cite{Arce2019}, a better correlation with thermal dust radiance is reported at degree-angular resolution, whereas the best correlation is found with WISE $12\,\si{\micro\meter}$ \citep{Wright2010WISE} at arcminute scales, which is interpreted as a hint that grains other than PAHs may dominate in the region. Regardless, \cite{Arce2019} raise the important caveat in spinning dust correlation analyses that the angular scale can play a major role. One possibility for why $\Re$ is a good template for $A_{\rm AME}$ in the region at degree-scales but not necessarily over the full sky is the assumption that $\tau_{353}$ is a proxy for $N_{\rm H}$, the hydrogen column density, where this relation generally depends on the environment \citep{Planck2014_XVII}. In the case of \mbox{$\lambda$ Orionis}, environmental variations may reduce the strength of the correlation between $A_{\rm AME}$ and $\tau_{353}$. The specific source of energy in the region (star \mbox{$\lambda$ Orionis}) also differs from the high-latitude sky, which is generally a softer integration over the local Galactic plane. In order to assess these relations over the entire sky, a component separation process such as \textsc{Commander} including low-frequency data such as \mbox{C-BASS} and \mbox{QUIJOTE} is necessary. This will enable the breaking of degeneracies between AME parameters and parameters from other emission mechanisms. This type of analysis is being implemented in the northern sky using \mbox{QUIJOTE} (Casaponsa et al. in prep.) and \mbox{C-BASS} data (Herman et al. in prep.).

The above discussion is reflected in the radiance- normalized emissivity, $A_{\rm AME}/\Re$, as a function of $G_0$, shown in plot \textit{L}, where no correlation is found with a coefficient of $0.29\pm0.10$. This is in part because $\Re$ and $A_{\rm AME}$ correlate well with each other to the extent that no measurable structure is left when taking the ratio between the two.  

Overall, the correlations between parameters shown in Fig.\,\ref{fig:correlations} point to a region where the ISRF from star \mbox{$\lambda$ Orionis} plays an important role in modifying the properties of the dust grains by a combination of dissociating grains and changing their size distribution, increasing their overall ionization and by providing an energy budget for higher peak frequencies closer to the star, with a combination of $\mathrm{EM}$ and $T_{\rm d}$ being the best predictors of $\nu_{\rm AME}$, $\Re$ being the best parameter to use as a tracer for the amplitude of AME in the region, and $G_0$ being a good predictor of the AME emissivity approximated as $A_{\rm AME}/\tau_{353}$.

\subsection{Are PAHs responsible?}
\label{sec:AKARI}

Determining the dominant carriers of spinning dust emission is a long-standing objective in AME research \citep{Dickinson2018AMEReview}. While it is likely that several species of AME carriers coexist due to the chemical diversity in dust grain populations, such as those labelled diffuse interstellar band (DIB) carriers \citep{Bernstein2015}, and that their relative abundances and contributions may be region-dependent, PAHs are generally considered a major candidate for diffuse spinning dust emission. This is due to the abundance and ubiquity of nanometer-sized PAHs with an electric dipole moment \citep{Tielens2008}.

In this section, we use the AKARI $9\,\si{\micro\meter}$ data in the region \citep{Bell2019}, shown in Fig.\,\ref{fig:multifrequency}, to assess correlations with the recovered AME amplitude map shown in Fig.\,\ref{fig:parameter_maps}, as well as correlations between $A_{\rm AME}$ and individual frequency maps in Table\,\ref{tab:surveys}. This is done through Spearman correlation coefficients calculated at \mbox{$N_{\rm side}=64$}, and simulated using the uncertainty distributions in $A_{\rm AME}$, which are a much greater contribution to the final uncertainties than noise in the individual frequency maps at \mbox{$N_{\rm side}=64$}. Calibration uncertainties do no directly affect the Spearman coefficients, since they are invariant under displacements such as changes in the absolute background level or scaling (e.g., calibration).

\begin{figure*}
	\begin{center}
		\includegraphics[width=\textwidth,angle=0]{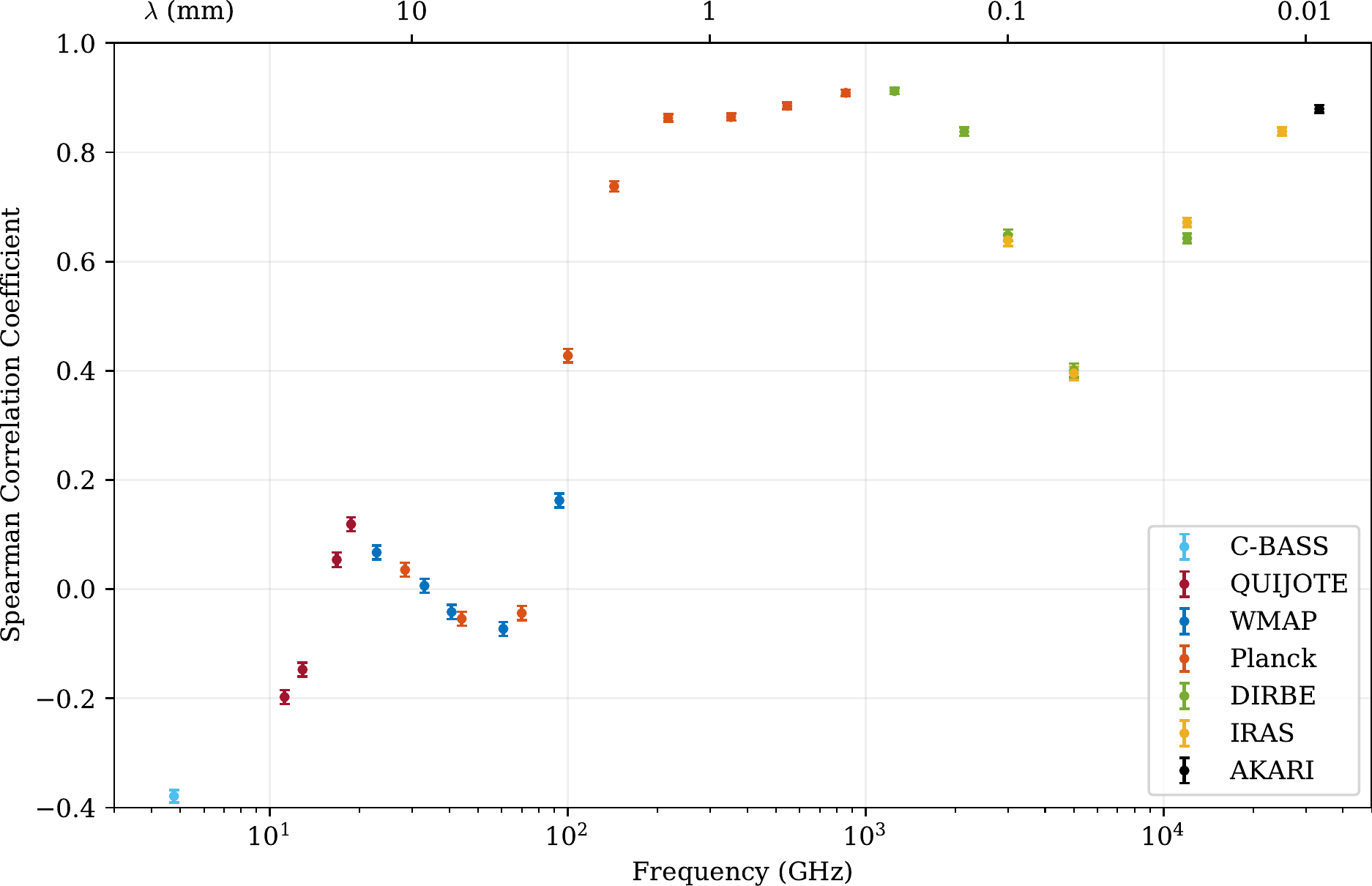}
		\caption{Spearman correlation coefficients between the AME amplitude map shown in Fig.\,\ref{fig:parameter_maps} and individual frequency maps. The main uncertainties come from the uncertainties in $A_{\rm AME}$.}
		\label{fig:spearman}
	\end{center}
\end{figure*}

The AKARI $9\,\si{\micro\meter}$ map in \mbox{$\lambda$ Orionis} was first reported to show a strong correlation with the \textsc{Commander} AME amplitude map by \cite{Bell2018}. This study is extended in \cite{Bell2019}, where correlations between the \textsc{Commander} maps and individual frequency maps are assessed. As discussed in Section\,\ref{sec:commander}, the \textsc{Commander} AME map suffers from degeneracies with free-free emission, so the \mbox{QUIJOTE} and \mbox{C-BASS}-aided $A_{\rm AME}$ map is a more reliable estimate of AME in the region. This is supported in this case by the fact that using \textsc{Commander} AME amplitudes reduces the absolute Spearman correlation coefficients at all frequencies relative to those calculated using our $A_{\rm AME}$ map. For example, at the DIRBE-COBE $1249\,\si{\giga\hertz}$ band, correlations drop from $0.912\pm0.006$ to $0.860$ when switching our AME map to the \cite{Planck2015_X} \textsc{Commander} separation. Similarly, correlations with the AKARI map drop from $0.879\pm0.007$ to $0.825$. The latter value is different from the coefficient of $0.89$ in \cite{Bell2019} due to the masking used in their study, which removes the brightest spots of emission, whereas we evaluate coefficients across the entire region including regions \textit{A} and \textit{B}.

The Spearman correlation coefficients as a function of frequency are shown visually in Fig.\,\ref{fig:spearman}, and numerically in Table\,\ref{tab:spearmantable}. Four main features can be seen: a peak around $\sim20\,\si{\giga\hertz}$, a range of correlations greater than 0.8 between $200<\nu<2000\,\si{\giga\hertz}$, a dip in the correlation coefficient near $60\,\si{\micro\meter}$, and an increase from that point up to $9\,\si{\micro\meter}$. The negative correlations below $\approx10\,\si{\giga\hertz}$ are due to the intrinsic spatial anticorrelation between AME in the region and the free-free emission in low-frequency maps. The coefficient peak at $\approx20\,\si{\giga\hertz}$ corresponds to the maximum AME contribution in the region relative to other emission mechanisms at those frequencies; primarily free-free emission, which results in overall coefficients at $\approx20\,\si{\giga\hertz}$ much lower than those at thermal dust frequencies. Spearman coefficients rise quickly above $100\,\si{\giga\hertz}$ due to the rapid overtaking of thermal dust emission as the dominant mechanism in CMB-subtracted maps. The highest correlations are reached at the \textit{Planck} $857\,\si{\giga\hertz}$ and DIRBE $1249\,\si{\giga\hertz}$ maps, with Spearman coefficients of $0.909\pm0.006$ and $0.912\pm0.006$, respectively. This coincides with the peak of thermal dust emission, and implies that these frequencies are the best tracers of AME in the region if the maps are used without any subtraction of background emission mechanisms, mainly due to the morphology of the underlying emission and in a small part due to their relatively high signal-to-noise ratio. As thermal dust emission from large grains starts to decrease above $\approx2000\,\si{\giga\hertz}$, sub-mm/IR emission from stochastically-heated grains out of local thermal equilibrium and contributions from very small grains (VSGs) become a significant component, which generally peaks around the minimum observed correlation near $60\,\si{\micro\meter}$, as shown in \mbox{\cite{Bell2019}}. In this region, transiently heated grains are not in thermal equilibrium, and there is excess emission near the center of the H\textsc{ii} region when compared with shorter wavelength maps such as $12\,\si{\micro\meter}$ mainly hotter dust. To a lesser extent, atomic lines such as the $63\,\si{\micro\meter}$ OI line can also add to the different morphology. The DIRBE and IRIS points give statistically consistent values where their frequencies overlap, with a small ($2.4\,\sigma$) discrepancy at $25\,\si{\micro\meter}$, likely from their different bandpass shapes. The rise in Spearman coefficients at $25\,\si{\micro\meter}$ is likely due to the overlap of the DIRBE and IRIS bands with the first PAH emission continuum. \cite{Smith2007} show typical measured PAH emission spectra in star forming galaxies, which can generally vary but give an idea of the possible spectra to be expected in \mbox{$\lambda$ Orionis}. The IRIS $12\,\si{\micro\meter}$ map also covers PAH features at 8.6, 11.2 and $12.7\,\si{\micro\meter}$ and is therefore expected to have a smaller relative contribution from VSGs \citep{Bell2019}, showing an even higher degree of correlation with $A_{\rm AME}$ than the $25\,\si{\micro\meter}$ map. The AKARI $9\,\si{\micro\meter}$ shows the highest correlation at near-IR frequencies with a Spearman coefficient of $0.879\pm0.007$. One possibility for this is the dominance of PAH emission in its band. The high degree of correlation implies that the dust and the PAHs are generally well-mixed in the region, as expected.

\begin{table}
\caption{Summary of the Spearman correlation coefficients shown visually in Fig. \ref{fig:correlations}, where the center of the band is given in either frequency or wavelength units according to the most typically used in each survey.  \label{tab:spearmantable}}
\resizebox{0.47\textwidth}{!}{
\begin{tabular}{@{}ccc@{}}
\toprule
\textbf{Instrument} & \textbf{Band Center}            & \textbf{Spearman Coefficient} \\ \midrule
C-BASS              & 4.76$\,$GHz              & $-0.379\pm0.011$         \\
QUIJOTE             & 11.2$\,$GHz              & $-0.197\pm0.013$         \\
QUIJOTE             & 12.9$\,$GHz              & $-0.147\pm0.013$         \\
QUIJOTE             & 16.8$\,$GHz              & $0.054\pm0.013$          \\
QUIJOTE             & 18.7$\,$GHz              & $0.119\pm0.013$          \\
WMAP                & 22.8$\,$GHz              & $0.067\pm0.013$          \\
Planck              & 28.4$\,$GHz              & $0.035\pm0.013$          \\
WMAP                & 33$\,$GHz                & $0.006\pm0.013$          \\
WMAP                & 40.7$\,$GHz              & $-0.042\pm0.013$         \\
Planck              & 44$\,$GHz                & $-0.054\pm0.013$         \\
WMAP                & 60.7$\,$GHz              & $-0.073\pm0.013$         \\
Planck              & 70$\,$GHz                & $-0.044\pm0.013$         \\
WMAP                & 93.6$\,$GHz              & $0.162\pm0.013$          \\
Planck              & 100$\,$GHz               & $0.428\pm0.012$          \\
Planck              & 143$\,$GHz               & $0.738\pm0.009$          \\
Planck              & 217$\,$GHz               & $0.863\pm0.007$          \\
Planck              & 353$\,$GHz               & $0.864\pm0.006$          \\
Planck              & 545$\,$GHz               & $0.885\pm0.006$          \\
Planck              & 857$\,$GHz               & $0.909\pm0.006$          \\
DIRBE               & $240\,\si{\micro\meter}$ & $0.912\pm0.006$          \\
DIRBE               & $140\,\si{\micro\meter}$ & $0.838\pm0.008$          \\
DIRBE               & $100\,\si{\micro\meter}$ & $0.648\pm0.011$          \\
IRAS                & $100\,\si{\micro\meter}$ & $0.639\pm0.011$          \\
DIRBE               & $60\,\si{\micro\meter}$  & $0.401\pm0.012$          \\
IRAS                & $60\,\si{\micro\meter}$  & $0.395\pm0.012$          \\
DIRBE               & $25\,\si{\micro\meter}$  & $0.642\pm0.009$          \\
IRAS                & $25\,\si{\micro\meter}$  & $0.672\pm0.009$          \\
IRAS                & $12\,\si{\micro\meter}$  & $0.838\pm0.008$          \\
AKARI               & $9\,\si{\micro\meter}$   & $0.879\pm0.007$          \\ \bottomrule
\end{tabular}}
\end{table}

The lower correlation in the AKARI $9\,\si{\micro\meter}$ is $3\,\sigma$ from the correlation coefficient for the \textit{Planck} $857\,\si{\giga\hertz}$ map near the thermal dust emission peak. However, this cannot be interpreted as evidence against PAH-dominated spinning dust in the region. The main reason for this is that residual emission uncorrelated with dust will reduce the Spearman coefficient, as can be seen in the case of the peak around $\approx20\,\si{\giga\hertz}$, in this case due to free-free emission. In addition, we use a different metric to trace the PAH column density by dividing the 9\,\si{\micro\meter} AKARI map by $G_0$, since the mid-IR emission mapped by AKARI depends on both column density of emitters and their temperature. In this case we find a lower correlation coefficient of $r_s=0.58\pm0.04$, primarily due to the relatively large uncertainties in the determination of $G_0$. The degree of dominance of thermal dust emission near its peak could therefore explain why the highest coefficients are obtained in this range, while AKARI has much higher relative residual diffuse contributions from VSGs and stochastically heated grains, which could account for the relatively small difference between the two coefficients. A secondary factor that can change the absolute level of Spearman correlations is the potential region-dependent abundance and contributions from different PAH species.

It is worth noting that while in the AME and thermal dust maps below $2\,\si{\tera\hertz}$ region \textit{A} is brighter than region \textit{B}, the reverse is true in both the AKARI and maps above $3\,\si{\tera\hertz}$. Since most information in measuring the Spearman correlations comes from the brightest spots (e.g., regions \textit{A} and \textit{B}), this reversal is likely responsible for the lower coefficients and is likely to come from relative PAH and VSG contributions.

While the subtraction of non-PAH mechanisms at near-IR bands is beyond the scope of this paper and our capabilities without additional near-IR data, two main conclusions can be drawn. Firstly, the reduction in correlations at sub-mm/IR frequencies is a hint that emission from stochastically-heated grains and VSGs has a different morphology to AME in the region. Secondly, the rise in coefficients at frequencies containing PAH bands above $60\,\si{\micro\meter}$ imply that PAHs are generally well-mixed with large grains and AME in the region. Overall, this highlights the need for follow-up high spatial and frequency resolution studies of AME regions, which have the potential to determine the dominant carrier molecules. 


\section{Conclusions}
\label{sec:conclusions}

This paper addresses three fundamental questions on low-frequency foregrounds, and specifically AME: Does the peak frequency of AME vary spatially with other environmental parameters, and how does this relate to current spinning dust models? How do PAH bands correlate with spinning dust and thermal dust emission, and could they be the dominant carriers? How reliable is the \cite{Planck2015_X} \textsc{Commander} AME separation, and how much do low-frequency data from \mbox{QUIJOTE} and \mbox{C-BASS} improve it in the region?

By extracting physical parameter maps of the region including an empirical log-normal approximation of the AME spectrum, we were able to make the first detection of variations in the AME peak frequency in a single region. The peak frequencies decrease radially outwards from star \mbox{$\lambda$ Orionis}, with individual radial profiles having a more well-defined functional form than the overall population. This is expected since the decrease in peak frequencies along each radii can depend on the local environment and the ring is only approximately circular. The spatial variation in peak frequencies is strongly correlated with the emission measures, dust temperatures and $G_0$, a proxy for the local radiation field. We also corroborated the previously reported result that the AME emissivity, approximated as $A_{\rm AME}/\tau_{353}$, is correlated with $G_0$ \citep{Tibbs2011,Tibbs2012,Planck2014_AME}. The correlation of $\nu_{\rm AME}$ and $A_{\rm AME}/\tau_{353}$ with the local radiation field is not expected in \textsc{SpDust} models since the models predict that the AME emissivity and peak frequency change very slowly over a large range of $G_0$ values, larger than those seen in the region. Therefore, this analysis is a hint towards the key role of the ISRF, primarily from \mbox{$\lambda$ Orionis}, on modifying the steady-state distribution of dust grain sizes and dipole moments in the region through a combination of effects. The changing of grain properties by the ISRF is an effect that \textsc{SpDust} currently does not directly model, since a single parameter can be changed independently while in practice changing one parameter can affect the others. The correlation between the AME and thermal dust radiances supports the idea that both emission mechanisms share the same power source in the region. The remarkable relations between the thermal dust radiance, $\Re$, and the $A_{\rm AME}$ amplitude make radiance a better predictor of AME in the region than $\tau_{353}$ due to the lower scatter and higher Spearman correlation coefficients in $\Re$.

This paper also assesses Spearman correlation coefficients between $A_{\rm AME}$ and individual frequency maps including the AKARI $9\,\si{\micro\meter}$ PAH-dominated map. The best frequency templates for the AME amplitude in the region is found to be the \textit{Planck} $857\,\si{\giga\hertz}$ and the DIRBE-COBE $1249\,\si{\giga\hertz}$ map, with all maps in the range $200<\nu<2000\,\si{\giga\hertz}$ exhibiting correlation coefficients above 0.8. A local minimum in correlation coefficients is found at the $60\,\si{\micro\meter}$ DIRBE and IRIS bands. The increase in correlation coefficients at 25, 12 and $9\,\si{\micro\meter}$, with increasing relative PAH contributions, suggests that PAHs in the region are well-mixed with AME carriers and large grains in thermal equilibrium.

Finally, by using \mbox{QUIJOTE} and \mbox{C-BASS}, we have shown that the \cite{Planck2015_X} \textsc{Commander} separation maps in the region suffer from degeneracies between free-free emission and AME due to the limited frequency coverage of the data used, highlighting the need for data below $\approx 20\,\si{\giga\hertz}$ as discussed in \cite{Genova-Santos2017} and \cite{Poidevin2019}. The addition of low frequencies emphasizes the importance of cross-collaborations at low frequencies such as \mbox{QUIJOTE} and \mbox{C-BASS}, where the overall outcome far exceeds the sum of their individual contributions. This is especially important for understanding AME at degree-scales.

While the \mbox{$\lambda$ Orionis} ring provides another piece of evidence for PDRs being good regions to look for AME and spatial variations in it, the main caveat of this paper is that the results above apply to this region at angular scales of a degree. Therefore, more regions and statistical studies are needed before these results can be extrapolated to the entire sky. Future work will aim to study variations in AME across regions at a higher resolution using instruments such as the Green Bank Telescope and the Sardinia Radio Telescope, together with mid-infrared observations of AME carrier candidates such as PAHs, fully exploring the nature of AME across PDRs and dark clouds. In order to confirm or rule out the dominance of different species as AME carriers, sub-mm/IR emission lines must also be disentangled, separating their contribution from the emission from other mechanisms. The highly-dimensional parameter space of spinning dust models makes linking theory to observations difficult, but with enough data a very interesting opportunity to study the properties and physics of the interstellar medium not traced by other emission mechanisms will arise. The improved constraining power of low-frequency data from experiments such as \mbox{QUIJOTE} and \mbox{C-BASS} should be used to build an improved component separation of AME, improving analyses such as \cite{Planck2014_AME} and breaking down the complexity of observations into fundamental relations with which the theory can be tested.


\section*{Acknowledgments}
The \mbox{QUIJOTE} experiment is being developed by the Instituto de Astrof\'{i}sica de Canarias (IAC), the Instituto de F\'{i}sica de Cantabria (IFCA), and the Universities of Cantabria, Manchester and Cambridge. The \mbox{C-BASS} project is a collaboration between Oxford and Manchester Universities in the UK, the California Institute of Technology in the U.S.A., Rhodes University, UKZN and the South African Radio Observatory in South Africa, and the King Abdulaziz City for Science and Technology (KACST) in Saudi Arabia. Partial financial support for \mbox{QUIJOTE} is provided by the Spanish Ministry of Economy and Competitiveness (MINECO) under the projects AYA2007-68058-C03-01, AYA2010-21766-C03-02, AYA2014-60438-P, AYA2017-84185-P, IACA13-3E-2336, IACA15-BE-3707 and EQC2018-004918-P; Agencia Estatal de Investigaci\'{o}n (AEI) and Fondo Europeo de Desarrollo Regional (FEDER, UE), under projects ESP2017-83921-C2-1-R and AYA2017-90675-REDC; the European Union's Horizon 2020 research and innovation programme under grant agreement number 687312 (RADIOFOREGROUNDS) and number 658499 (PolAME); Unidad de Excelencia Mar{\'\i}a de Maeztu (MDM-2017-0765), and the Consolider-Ingenio project CSD2010-00064 (EPI: Exploring the Physics of Inflation). \mbox{C-BASS} has been supported by the NSF awards AST-0607857, AST-1010024, AST-1212217, and AST-1616227, and NASA award NNX15AF06G, the University of Oxford, the Royal Society, STFC, and the other participating institutions. \mbox{C-BASS} is also supported by the South African Radio Astronomy Observatory, which is a facility of the National Research Foundation, an agency of the Department of Science and Technology. RCA would like to thank Aaron Bell, Takashi Onaka and the AKARI collaboration for providing access to the AKARI $9\,\si{\micro\meter}$ data in the region. We make use of the \texttt{HEALPix} package \citep{Gorski2005} and Python \texttt{astropy} \citep{astropy2013,astropy2018}, \texttt{corner}, \texttt{emcee} \citep{Foreman-Mackey2013}, \texttt{healpy} \citep{healpy}, \texttt{matplotlib} \citep{matplotlib}, \texttt{numpy} \citep{numpy} and \texttt{scipy} \citep{SciPy} packages. RCA acknowledges support from an STFC postgraduate studentship and a President's Doctoral Scholar Award from the University of Manchester. FP acknowledges support from the Spanish Ministerio de Ciencia, Innovaci\'{o}n y Universidades (MICINN) under grant numbers ESP2015-65597-C4-4-R, ESP2017-86852-C4-2-R, ESP2015-65597-C4-4-R and ESP2017- 86852-C4-2-R. The authors thank the referee for useful feedback and recommendations.

\section*{Data Availability}
The data presented in this article can be made available on request to the author or the \mbox{C-BASS} or \mbox{QUIJOTE} collaborations. In the near future, both \mbox{C-BASS} and \mbox{QUIJOTE} data will be publicly released.


\bibliographystyle{mn2e}
\bibliography{paper_refs}

\label{lastpage}

\end{document}